\documentclass[aps,prd,floatfix,prd,showpacs,superscriptaddress,preprintnumbers,twocolumn,nofootinbib]{revtex4-1}
\usepackage{graphicx}
\usepackage{dcolumn}
\usepackage{bm,latexsym,amsmath,amssymb,amsfonts}
\usepackage[colorlinks]{hyperref}
\usepackage{color}
\usepackage{gensymb}
\usepackage{adjustbox}
\usepackage{scrextend}
\usepackage{enumerate}
\usepackage{aas_macros}

\usepackage{mwe}

\usepackage{natbib}
\usepackage{enumitem,kantlipsum}

\def\apjs{Astrophys.~J. Suppl.}

\def\apjl{Astrophys.~J.~Lett.}

\def\mnras{Mon.~Not.~Roy.~Astr.~Soc.}

\newcommand{\be}{\begin{equation}}
\newcommand{\ee}{\end{equation}}

\begin{document}

\title{Improved Binary Pulsar Constraints \\ on the Parameterized post-Einsteinian Framework}
\author{Remya Nair}
\email{remya.nair@montana.edu}
\affiliation{eXtreme Gravity Institute, Department of Physics, Montana State University, Bozeman, MT 59717, USA}

\author{Nicol\'as Yunes}
\email{nyunes@illinois.edu}

\affiliation{eXtreme Gravity Institute, Department of Physics, Montana State University, Bozeman, MT 59717, USA}
\affiliation{Department of Physics, University of Illinois at Urbana-Champaign, Urbana, IL 61801, USA}
\date{\today}

\begin{abstract}
The parameterized post-Einsteinian formalism was developed to search for generic deviations from general relativity with gravitational waves. 
We here present constraints on this framework using Bayesian analysis of a set of binary pulsar observations. 
In particular, we use measurements of the Keplerian and post-Keplerian parameters of six different binary pulsar systems, and Markov-Chain Monte-Carlo exploration to calculate posteriors on the parameterized post-Einsteinian parameters and derive robust constraints. 
We find improvements of 1--2 orders of magnitude in the strength of constraints when combining all six observations, relative to what one can achieve when using only the double binary pulsar.
We also find that the constraints are robust to any correlation with the binary's component masses. 
The bounds on the parameterized post-Einsteinian framework derived here could be used as a prior in future Bayesian tests of general relativity with gravitational wave observations. 
\end{abstract}

\maketitle

\section{Introduction}\label{intro}

\begin{figure*}[ht]
\includegraphics[width=0.475\textwidth]{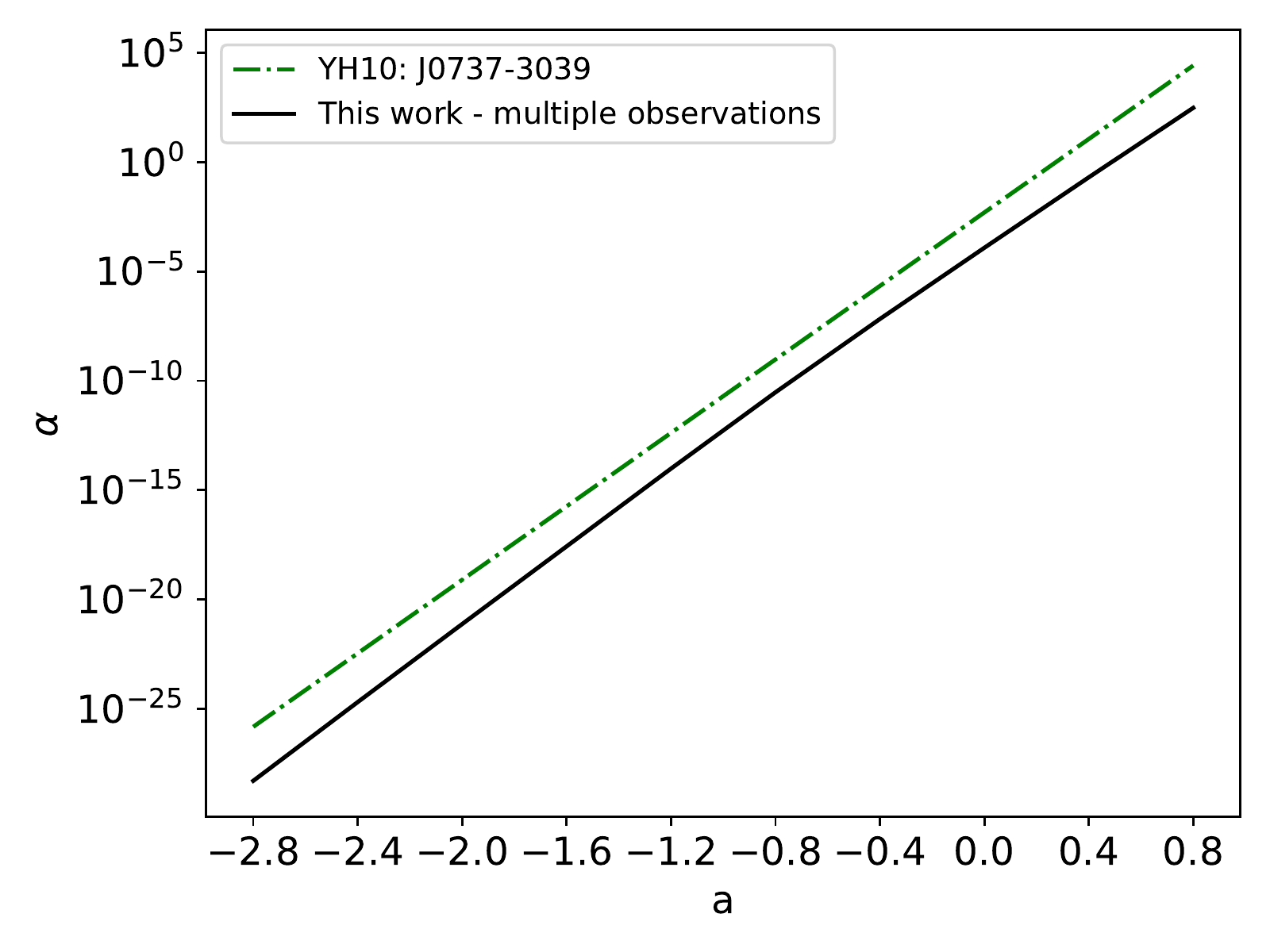}
\includegraphics[width=0.475\textwidth]{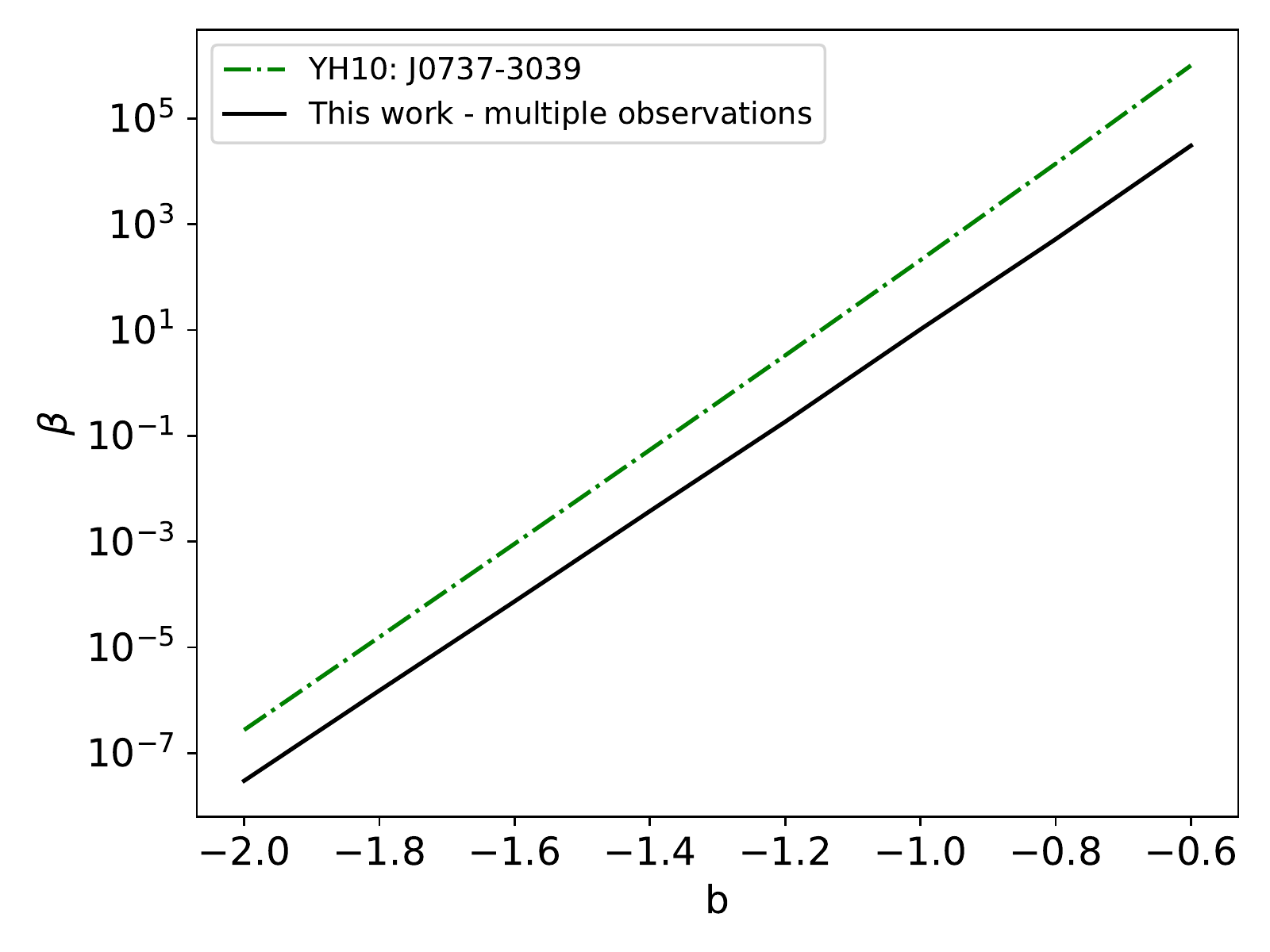} 
	\caption{Joint 95 percentile upper limits on the ppE amplitude parameter $\alpha$ (left) and ppE phase parameter $\beta$ (right) as a function of the ppE amplitude exponent $a$ (left) and $b$ (right), after marginalizing over the component masses, from the observation of six binary pulsars: J2222-0137, J1012-5307, J0348-0432, J0737-3039, J1909-3744 and J1738-0333 \cite{Kramer:2006nb,Kaplan:2014mka,Cognard:2017xyr,Lazaridis:2009kq,1998MNRAS.298..207C,Antoniadis:2013pzd,Reardon:2015kba,Jacoby:2003nq,Antoniadis:2012vy,Freire:2012mg}. These limits are obtained using Gaussian mass priors informed from GR estimates (see \ref{priors: gr} in Sec.~\ref{sec:priors}). The region above the (black) solid line is ruled out by these observations. For comparison, we also include here the relational constraints derived in~\cite{Yunes:2010qb}, which are 1--2 orders of magnitude weaker than those obtained here. }
	\label{fig:ppE_const_GR}
\end{figure*}


Einstein's theory of gravity, General Relativity (GR), provides us with the essential tools to study the large scale structure of our Universe. Since the advent of GR, the field of cosmology has steadily been on the path to becoming a high precision science. With the spectacular measurements of the cosmic microwave background anisotropy by WMAP~\cite{Bennett:2012zja} and later Planck~\cite{Aghanim:2018eyx}, we are indeed living in an era of precision cosmology. But, our standard theoretical paradigm for the Universe, the inflationary Big Bang model, is still incomplete, as the physics behind the late time acceleration of the Universe remains elusive and the initial conditions that led to inflation (accelerated expansion in the early Universe) are still unknown. The standard flat-$\Lambda$CDM model is also currently under scrutiny from the latest Planck findings that prefer a positive curvature at more than 99\% confidence level \cite{DiValentino:2019qzk}. All of this has suggested to some that a more robust description of the accelerating Universe is perhaps conceivable through a modification of GR. 

Einstein's theory, however, has passed a plethora of tests with flying colors, from observations in the Solar system that explore the quasi-stationary weak field regime~\cite{Will:2014kxa}, to binary pulsar observations that probe gravity in the (quasi-stationary) strong field regime~\cite{Lebach:1995zz,Everitt:2011hp, Kramer:2012zz}. The latter are particularly constraining since they can be sensitive to gravity modifications that are suppressed in the Solar system. In particular, binary pulsar observations are affected by the back reaction of gravitational waves (GWs) on the orbital dynamics of binary systems~\cite{Hulse:1974eb}, which is simply not accessible in the Solar system. The observation of such orbital decay requires extremely precise timing of the radio `pulses' that originate from the magnetic poles of the pulsars~\cite{Hewish:1968bj}. Given a set of measurements of arrival times, radio astronomers can then fit a timing model that can extract the (Keplerian) orbital parameters of the binary system. If the observed pulsar is sufficiently stable (as is the case for milli-second pulsars), then the timing model can typically fit every pulse over a time span of many months or even several years, allowing for the extraction of post-Keplerian parameters, like the orbital period decay. 

The discovery of binary pulsars provided the perfect ground to test the strong field regime of the gravitational interaction. In a series of papers, Damour and Deruelle developed the `parameterized post Keplerian' (ppK) framework to study the quasi-stationary regime of binary pulsars, and they parameterized all the observables obtained from pulsar timing
\cite{Damour:1985d, Damour:1985db}; this was later further developed by Damour and Taylor in the context of tests of modified gravity theories \cite{Damour:1991rd}. Given a theory of gravity, one can relate the ppK parameters to the Keplerian orbital parameters and the component masses of the binary system. Since the component masses are the only unknowns in the system that are not observed directly, it follows that a measurement of any two ppK parameters uniquely determines the two masses, and the values for other
ppK parameters can be predicted. If any other ppK parameter is measured, it provides a consistency test for the underlying theory of gravitation. Hence, the ppK framework allows for  theory-independent tests of gravity using binary pulsar observations. 

Let us consider the case of the double-pulsar J0737-3039A/B to appreciate the power of pulsar timing. This binary system was discovered in 2003 and soon enough a total of six ppK parameters were measured for the system. The relativistic precession of the orbit was known to better than 0.004\%, back in 2006 \cite{Kramer:2006nb} and, consistent measurements would further decrease the uncertainties. All the ppK parameters are consistent with GR and this double-pulsar provides the best test for the GR quadrupole formalism for GW generation \cite{Kramer:2012zz}. Other pulsar systems have been used to test the validity of various assumptions of GR, one of which for example, is the absence of dipolar emission. 
Binary pulsar systems with significantly different compactness (the ratio of their mass to their radius) can be used for testing dipolar radiation. In the double pulsar system, the compactness are similar, since both objects are NSs, and hence dipolar radiation (if it exists) is strongly suppressed. On the other hand, PSR J1738+0333 is composed of a low mass white dwarf and a pulsar companion, so the compactness are wildly different and this system can be used to stringently constrain modified gravity theories that predict the existence of dipolar emission~\cite{Gerard:2001fm}.

Binary pulsar observations, however, are no longer alone as probes of GR in the strong field. The latest additions to gravity probes are GWs from coalescing binaries, which help us inspect the highly-dynamical, strongly-curved spacetimes around massive objects. As with binary pulsars, a generic framework to perform model independent tests of GR with GWs was developed in the last decade~\cite{Yunes:2009ke}. The parameterized post-Einsteinian (ppE) formalism proposes to augment the GR predictions for the GW models with additional (non-GR) parameters. GR and other modified theory predictions, such as scalar tensor theories~\cite{1992CQGra...9.2093D}, dynamical Chern-Simons gravity~\cite{Alexander:2009tp}, Einstein-dilaton Gauss-Bonnet gravity~\cite{Kanti:1995vq}, among others~\cite{Yunes:2013dva}, can then be recovered by specific choices of these non-GR parameters. The LIGO-VIRGO collaboration has observed GWs from multiple compact binary mergers, and put some constraints on the allowed ppE deviations from GR~\cite{TheLIGOScientific:2016src, Abbott:2017vtc, Abbott:2018lct, LIGOScientific:2019fpa}. These ppE constraints can be mapped to constraints on specific modified gravity theories, as was done in \cite{Yunes:2016jcc, Nair:2019iur}. As the number of these events increases, we can expect the constraints to become more and more stringent~\cite{LIGOScientific:2019fpa}. 

The ppE modifications to the GWs emitted by binary systems do not just affect the observables of GW detectors, but they also modify ppK parameters. For example, ppE modifications to the GW amplitude or to the GW phase affect the rate of decay of the orbital period of any binary system, since the latter depends on the energy flux carried away from the system by GWs. In \cite{Yunes:2010qb} (from here on YH10), the authors showed explicitly how ppE modifications affect the rate of orbital period decay, and then they derived relational constraints on the ppE parameters as a function of the post-Newtonian (PN) order\footnote{The PN approximation is one in which the field equations are solved as an expansion about small velocities and weak fields. An expansion of $N$PN order is one that is proportional to $v^{2N}$ relative to its leading-order, controlling factor~\cite{Blanchet:2013haa}.} at which they appear, given binary pulsar observations. These relational constraints were then evaluated using measurements of the orbital decay of the double binary pulsar PSR J0737-3039. To obtain these constraints, the component masses were fixed to their best fit values, obtained by radio astronomers from the observations of Keplerian and post-Keplerian parameters of the binary while assuming GR is valid. 

In this paper, we extend the study of YH10 by carrying out a Bayesian analysis on six binary pulsar observations to derive posteriors on the ppE parameters, marginalized over the component masses. One of our main results is shown in Fig. \ref{fig:ppE_const_GR}, where we plot the 95 percentile upper limit on the amplitude and phase ppE corrections, as a function of the PN order at which these corrections appear (see Eq. \eqref{eq: h_ppE}). These limits are obtained using Gaussian priors on the component masses (see \ref{priors: gr} in Sec.~\ref{sec:priors}), informed from the GR estimates which in turn are obtained from fitting pulsar timing data to a GR timing model. As compared to the results of YH10, also plotted in this figure, we see that our bounds are better by 1--2 orders of magnitude at all PN orders. This is because we use more observations than in YH10, and the improvement in the constraints are in spite of carrying out a Bayesian analysis in which the components are allowed to vary. These constraints are thus more stringent and more robust than those obtained previously, and should be used as informative priors for future studies with binary pulsars or GW observations. 

We want to make an important note here. In combining multiple binary observations and
	obtaining these constraints, we have made an implicit assumption that the non-GR correction is independent of the system parameters like masses and spins. If this is not the case, then combining measurements from different binary systems, with different masses/spins (and hence different values for the non-GR corrections) will not be justified. In this scenario one has to carry out a theory specific analysis, by incorporating the exact dependence of the non-GR corrections on the system parameters.

The remainder of this paper explains the details involved in obtaining these results and it is organized as follows. In Section~\ref{ppE}, we present a brief overview of parameterized tests of GR, and how they can be used to obtain evidence or constrain deviations from GR in a model independent manner. In Section~\ref{bayesian}, we discuss our Bayesian scheme and how we obtain our upper limits on the ppE parameters. In Section~\ref{results}, we conclude with a discussion of our results and implications. Henceforth, we used geometric units in which $G  = 1 = c$ unless otherwise stated.

\section{Parameterized tests of GR}
\label{ppE}

In this section, we introduce the ppE and the ppK frameworks and show how they lead to modifications to binary pulsar observables. The interested reader is referred to~\cite{Yunes:2009ke, Yunes:2013dva} and~ \cite{Damour:1985d, Damour:1985db, Damour:1991rd} for more details on these frameworks.

\subsection{ppE framework}
The ppE framework was introduced by Yunes and Pretorius to study deviations from Einstein equations, in a systematic and model-independent manner \cite{Yunes:2009ke}. In this sense, the ppE framework is similar to the parameterized post-Newtonian (ppN) formalism, which was built as an expansion about Minkowski space to describe weak field interactions in the Solar system, or the ppK framework mentioned in the previous section. The advantage of the ppE framework is that one can measure or constrain generic deviations from GR predictions instead of looking for some specific kind of deviation that may be predicted by a particular (alternative) theory.

The authors of~\cite{Yunes:2009ke} focused on GWs emitted by the quasi-circular inspiral, merger, and ringdown of binary black holes. The deviations were encoded in the GW model of the detector response to an impinging GW through corrections in both the GW amplitude and the GW phase in the frequency domain. The standard ppE framework proposes modifications to the frequency-domain waveform corresponding to the two GR polarizations, but since then, the framework has been expanded to allow for the presence of additional polarizations~\cite{PhysRevD.86.022004}. The frequency-domain corrections arise due to modifications in the binding energy of the binary system or its rate of change. The proposed enhancements correspond to deviations in the waveform amplitude and phase, which can be written as
\begin{equation}
\tilde{h}(f) = \tilde{h}_{\rm GR} (1+\alpha u^a) \exp^{i\beta u^b}, ~~u = (\pi {\cal M}f)^{1/3},
\label{eq: h_ppE}
\end{equation}
where $\tilde{h}(f)$ is the ppE modified (Fourier-domain) waveform, $\tilde{h}_{\rm GR}$ is the GR waveform, ${\cal M} = (m_1m_2)^{3/5}/(m_1+m_2)^{1/5}$ is the chirp mass, with component masses $m_{1,2}$, and $f$ is the GW frequency. We see that in addition to the system parameters, like the masses, the modified waveform also depends on ppE parameters. The ppE parameters $\alpha$ and $a$ are amplitude parameters, while $\beta$ and $b$ are ppE phase parameters. 

The ppE parameters $\alpha$ and $\beta$ control the magnitude of the GR deviations, and thus, they can be constrained from observations, while $a$ and $b$ determine the PN order at which the modification enters the waveform, and hence they characterize the type of physical modification. Note that $b=(k-5)/3$ corresponds to correction appearing at the (k/2)PN order. A correction with $b=-5/3$ corresponds to a modification at leading PN order (0PN), and hence, corrections entering with $b<-5/3$ and $b>-5/3$ are understood as {\emph negative} and {\emph positive} PN corrections respectively. In general, $\alpha$ and $\beta$ can depend on the system parameters, in addition to any fundamental (coupling) constants introduced by the alternative theory. GR predictions can be recovered in this framework by simply setting ($\alpha, \beta$) =(0,0). 

The ppE waveforms can be thought of as a result to generic modifications to the orbital chirping rate of the binary, $\dot{F}$ (= $\dot{E}(dE_b/dF)^{-1}$). This rate can be modified if one modifies the conservative sector i.e.~the orbital binding energy $E_{b}$, or if one modifies the dissipative sector i.e.~the amount of energy lost from the binary systems due to emission of GW (or any other propagating degree of freedom), $\dot{E}$. Since both modifications lead to changes in the GW frequency and phase, the detection of a deviation from GR does not allow one to determine whether this deviation was a result of modifications to the conservative or the dissipative sector. 

\subsection{ppK framework}
\label{ppK_binPuls}

As discussed briefly in Sec.~\ref{intro}, in order to study the strong field regime of binary pulsar observations, and the back reaction on the orbital dynamics due to GW, one can use the ppK parameterization. This provides a way to obtain theory independent information by fitting the Keplerian and post-Keplerian parameters in a timing model. There are five Keplerian parameters that are usually employed to describe the orbital dynamics: the orbital period $P_b$, the orbital eccentricity $e$, the projected semi-major axis $x$, the longitude of periastron $\omega$, and the time of periastron passage $T_0$.  For any gravity theory, the post Keplerian parameters can be written as functions of one or more of these Keplerian parameters, the component masses and any extra parameters (new fundamental constant) that the theory may introduce. 

Consider the post-Keplerian parameter $\dot{P_b}$ which is related to the orbital period damping due to GW emission.  In GR, one can relate this post-Keplerian parameter to the Keplerian parameters $P_b$ and $e$ via \cite{Peters:1964zz}
\begin{equation}
\dot{P_b}^{\rm GR} = - \frac{192 \pi}{5} \frac{m_p m_c}{M^2} \frac{1+73 e^2/24 + 37 e^4/96}{(1-e^2)^{7/2}} \frac{V_b^5}{c^5}.
\label{eq:pbdot_GR}
\end{equation}
where $m_p$ and $m_c$ are the pulsar and companion masses respectively, $M$ is the total mass and $V_b \equiv  (GMn_b)^{1/3}$, with $n_b \equiv 2\pi/P_b$. Observe that this post-Keplerian parameter (and in fact other post-Keplerian parameters in GR) are independent of the internal structure of the component objects, which may not be true in modified theories. For example, in mono-scalar-tensor theory, the rate of change of the orbital period depends on the difference between the effective scalar coupling $\alpha$ of the component objects \cite{Will:1977wq,Damour:1996ke} via
\begin{equation}
\dot{P_b}^{\rm ST} = - 2 \pi \frac{m_p m_c}{M^2} \frac{1+ e^2/2}{(1-e^2)^{5/2}} \frac{{\cal V}_b^3}{c^3} \frac{(\alpha_p - \alpha_c)^2}{1+\alpha_p \alpha_c} + {\cal O}({\cal V}_b^5/c^5),
\label{eq:pbdot_ST}
\end{equation}
where the subscripts $p$ and $c$ stand for the pulsar and its companion, and ${\cal V}_b = (G_*(1+\alpha_p \alpha_c)Mn_b)^{1/3}$.

Other post-Keplerian parameters include the change in the longitude of periastron $\dot{\omega}$, change in the orbital eccentricity and the projected semi-major axis, $\dot{e}$ and $\dot{x}$. Besides these, there are  post-Keplerian parameters which are important due to their relativistic effect that alter the timing of the arrival of the radio pulses: Einstein delay $\gamma$ and the Shapiro time delay $r,s$. In this work we focus only on $\dot{P_b}$, since this is the only parameter that has been related to the ppE corrections.

\begin{flushright}
	
\end{flushright}
\subsection{Mapping between ppK and ppE frameworks}
\label{map_pp}

As in the $\dot{P_b}$ case, other post-Keplerian parameters can also be related to Keplerian parameters, and then fitted with binary pulsar measurements, but for the purpose of this paper, we will focus on the orbital period decay ${\dot{P_b}}$. As we will show next, we can relate the post-Keplerian parameter ${\dot{P_b}}$ to the post-Einsteinian parameters $\alpha$ and $\beta$ at any fixed PN order determined by the post-Einsteinian parameters $a$ and $b$. Such a mapping is not so simple (and in fact, has not yet been derived) for other ppK parameters, because the ppE formalism has not yet been extended to eccentric orbits. 

In order to put constraints on post-Einsteinian parameters using binary pulsar measurements, we first need to map the ppK parameters, in the present case $\dot{P_b}$, to the ppE corrections we introduced in Eq.~\eqref{eq: h_ppE}. If the binding energy of the binary system is the same as in GR, the gravitational wave luminosity $\dot{E}$ can be related to the orbital decay $\dot{P}_{b}$ as: 
\begin{equation}
\frac{\dot{P}_{b}}{P_{b}} = \frac{3}{2} \frac{\dot{E_b}}{E_b} = - \frac{3}{2} \frac{\dot{E}}{E_b},
\label{eq:energy_b}
\end{equation}
where $\dot{E}$ is the energy lost due to GW emission (and the emission of any additional propagating degree of freedom), $E_b$ is the binding energy of the system and in the second equality we have used energy balance, i.e., the amount of binding energy lost due to GW emission is equal to (negative of) the amount of energy carried away by GWs (and any other propagating field). 

As discussed in Sec.~\ref{ppE}, the ppE corrections cannot distinguish between modifications due to corrections to the binding energy or to the GW luminosity (or to both).  In YH10, the authors considered modifications in the dissipative sector, i.e. in the energy carried away by the GW, while keeping the binding energy the same as in GR. Under this assumption, they found that $\dot{E}$ is modified as follows
\begin{equation}
\dot{E} = \dot{E}_{\rm GR} \left[1+ \pi^2 {\cal M}^2 \beta ~b (b-1) u^{b-2} \left(\frac{d^2 \Psi_{\rm GR}}{df^2} \right)^{-1} \right],
\label{eq: dotE}
\end{equation}
where $\Psi_{\rm GR}$ is the GW phase in GR. The quantity $\dot{E}_{\rm GR}$ for a binary system in an eccentric orbit is~\cite{Peters:1963ux} 
\begin{equation}
\dot{E}_{\rm GR} = - \frac{32}{5} \eta^2 \frac{M^5}{r_{\rm 12}^5} (1-e^2)^{-7/2} \left(1+\frac{73}{24} e^2 +\frac{37}{96} e^4\right),
\end{equation}
where $r_{\rm 12}$ is the length of the semi-major axis. Using these expressions, and in the light of the energy balance equation [Eq.~\eqref{eq:energy_b}], one can write the phase corrected orbital decay as
\begin{equation}
\frac{\dot{P_b}}{P_b} = \left( \frac{\dot{P_b}}{P_b} \right)_{\rm GR} \left(1+ \frac{48}{5} \beta b(b-1) u^{b+5/3} \right), \label{eq:decay_phase}
\end{equation}
thus relating the post-Keplerian parameter $\dot{P}_{b}$ to the post-Einsteinian parameter $\beta$ at fixed $b$. Note that one can do a similar exercise of relating the ppE phase correction to $\dot{P}_{b}$ assuming that the correction comes from the conservative sector (changes in the binding energy of the orbit) as shown in \cite{PhysRevD.88.064056}.

The GW amplitude also depends on the rate of change of the orbital frequency, $\dot{F}$, and so we can also express the GW luminosity in terms of the amplitude ppE parameters. YH10 followed this logic and found that $\dot{E}$ can be expressed in terms of amplitude ppE parameters as
\begin{equation}
\dot{E} = \dot{E}_{\rm GR}  (1+ \alpha u^a)^2.
\end{equation}
This in turn can be used to obtain the amplitude corrected orbital decay as:
\begin{equation}
\frac{\dot{P_b}}{P_b} = \left( \frac{\dot{P_b}}{P_b}\right)_{\rm GR} \left(1+ 2\alpha u^a \right).
\label{eq:decay_amp}
\end{equation}
where the GR prediction is
\begin{equation}
\left(\frac{\dot{P_b}}{P_b}\right)_{\rm GR} = - \frac{96}{5} \frac{\eta m^3}{r_{12}^4} (1-e^2)^{-7/2} \left(1+\frac{73}{24} e^2 + \frac{37}{96} e^4 \right).
\end{equation}
As noted in YH10, observe that there is a one to one mapping between the amplitude and phase correction parameters, namely $a=b+5/3$ and $\beta = 5 \alpha/(48b(b-1))$ if the waveform amplitude and phase are modified due to the same mechanism. 

\begin{table*}[ht]
	\centering
	\begin{tabular}{l c c c c c}
		\hline
		\rule{0pt}{4ex} ~~System & $P_b$[days] & $e$ & $\dot{P_b}$ & $\dot{P_b}/P_b$ & Ref.\\
		\hline
		\hline
		\rule{0pt}{4ex} J2222-0137 & 2.44576469(13) & $3.8 \times 10^{-4}$ & $-0.06(9) \times 10^{-12}$ & $-2.839 \pm 4.259 \times 10^{-19}$ & \cite{Kaplan:2014mka,Cognard:2017xyr}\\
		\hline
		\rule{0pt}{4ex} J1012-5307 & 0.60467271355(3) & $1.2 \times 10^{-6}$ & $-0.15 (15) \times 10^{-15}$& $-2.871 \pm 2.871 \times 10^{-19}$  & \cite{Lazaridis:2009kq,1998MNRAS.298..207C}\\
		\hline
		\rule{0pt}{4ex} J0348-0432 & 0.102424062722(7) & $2.6 \times 10^{-6}$ & $-2.74 (45) \times 10^{-13}$ & $-3.096 \pm 0.508 \times 10^{-17}$ & \cite{Antoniadis:2013pzd}\\
		\hline
		\rule{0pt}{4ex} J0737-3039 & 0.10225156248(5) & $8.7 \times 10^{-2}$ & $-1.25 (17) \times 10^{-12}$& $-1.417 \pm 0.019 \times 10^{-16}$  & \cite{Kramer:2006nb}\\
		\hline
		\rule{0pt}{4ex} J1909-3744 & 1.533449474406(13) & $1.1 \times 10^{-7}$ & $-6 (15) \times 10^{-15}$ & $-4.528 \pm 0.132 \times 10^{-20}$  & \cite{Reardon:2015kba,Jacoby:2003nq}\\
		\hline
		\rule{0pt}{4ex}  J1738-0333 & 0.3547907398724(13) & $3.5 \times 10^{-7}$ & $-25.9 (3.2) \times 10^{-15}$ & $-8.449 \pm 1.043 \times 10^{-19}$ & \cite{Antoniadis:2012vy,Freire:2012mg}\\
		\hline
	\end{tabular}
	\caption{
		Keplerian and post-Keplerian parameters for the six binary pulsar systems used in this work. The figures in parentheses represent the estimated $1\sigma$-uncertainties in the last quoted digit. The eccentricities of all of these systems are very well measured, do not contribute significantly to the error budget and, are therefore ignored in our analysis. For easy reference we also show the estimated values and 1$\sigma$ uncertainty on the variable  ($\dot{P_b}/P_b$). Interested readers can refer to the cited papers to find the values of Keplerian and post-Keplerian measurements for these binaries.
	}
	\label{tab:data}
\end{table*}

Equations \eqref{eq:decay_phase} and \eqref{eq:decay_amp} show us how to use the measurements of post-Keplerian parameters obtained from binary pulsar observations to constrain post-Einsteinian parameters. In YH10, the authors obtained \emph{relational constraints} on the magnitude of the ppE corrections ($\alpha$ and $\beta$) by relating them to the binary pulsar observational error $\delta$, which is indicative of the accuracy with which the orbital decay is measured. To be clear, the authors of YH10 assumed that the observed ppK parameter $(\dot{P}_{b})_{\rm o}$ matched its predicted value in GR  $(\dot{P}_{b})_{\rm th}$, up to some uncertainty $\delta$ (in units of Hz), i.e.
\be
\label{eq:relational}
\left( \frac{\dot{P}_{b}}{P_{b}}\right)_{\rm o} = \left( \frac{\dot{P}_{b}}{P_{b}}\right)_{\rm GR} + \delta\,. 
\ee
Then, they set this equation equal to either Eq.~\eqref{eq:decay_phase} or~\eqref{eq:decay_amp}, and solved for $\alpha$ and $\beta$ to find a constraint in terms of Keplerian parameters and the uncertainty $\delta$.  Upper limits on the post-Keplerian parameters can then be obtained for different values of $a$ and $b$ (corresponding to deviations arising at different PN orders) assuming a value for the Keplerian parameters, which YH10 took to be the best-fit values obtained from the fits to a GR timing model.  

\section{Bayesian analysis and parameter estimation} 
\label{bayesian}
In this section we describe the data set used in this work and give a brief overview of the Bayesian formalism. 
In YH10, relational constraints were obtained on the ppE amplitude and phase parameters, by fixing the component masses to the values predicted by assuming the validity of GR. Although this is a good approximation on the upper limits of these parameters, a more robust method to obtain constraints is to do a joint Bayesian parameter estimation study varying over all free parameters in the model, i.e.~over the post-Einsteinian parameter \emph{and} the component masses simultaneously. Note that we do not allow the eccentricity to vary because for the binary pulsars considered here, the orbits are all very nearly circular. We use a Bayesian framework for parameter estimation and study the effect of various mass priors on the estimation of the ppE corrections, considering the observations of multiple binary pulsars. We expect that the inclusion of more binary pulsar observations will enhance the constraints on ppE parameters, while allowing the masses to vary may have the opposite effect.  In this section, we give a brief overview of the data used, the Bayesian scheme and the Markov Chain Monte-Carlo (MCMC) sampling approach we will employ to estimate the posterior distribution of masses and ppE corrections.

\subsection{Data} 
We use six binary pulsar observations of the orbital decay to put constraints on the amplitude and phase corrections appearing at various PN orders. These are J2222-0137, J1012-5307, J0348-0432, J0737-3039, J1909-3744 and J1738-0333 \cite{Kramer:2006nb,Kaplan:2014mka,Cognard:2017xyr,Lazaridis:2009kq,1998MNRAS.298..207C, Antoniadis:2013pzd,Reardon:2015kba,Jacoby:2003nq,Antoniadis:2012vy,Freire:2012mg}. All of these binaries have low eccentricities ($e<0.1$) and consist of a pulsar with either a NS or a white dwarf companion. In order to obtain constraints on the post-Einsteinian parameter $\alpha$ or $\beta$ using Eq.~\eqref{eq:decay_phase} and \eqref{eq:decay_amp}, we work with the measurements of the Keplerian parameters $\left\lbrace P_b,e \right\rbrace$ and the post-Keplerian parameter $\dot{P_b}$; the best-fit measurements and the uncertainties of these quantities for the different binary pulsars we considered, are presented in Table \ref{tab:data}. The post-Einsteinian parameter $\alpha$ or $\beta$ and the component masses will be free parameters of the model (more discussion to follow). 

Before proceeding, let us make a clarifying point here. Recall that in GR, the measurement of the two masses in the binary requires at least $2$ ppK measurements, such as $\dot{P}_{b}$ and the Einstein time delay parameter $\gamma$. What would happen if we only used one ppK measurement, such as $\dot{P}_{b}$? If so, there would not be enough information in the data to break the $\{m_{p},m_{c}\}$ degeneracy, and we would not be able to independently constrain the two masses. Therefore, one should expect that any attempt to constrain ppE amplitude parameters and the masses of the binary using only $\dot{P}_{b}$ should fail. As we will show later, this is not quite the case: even when one assumes nothing about the masses, one can still constrain the ppE amplitude parameters, albeit less strongly than when one assumes the masses are known to some degree. The masses themselves, however, are indeed very poorly estimated when one uses only $\dot{P}_{b}$ and assumes no mass priors.   

\subsection{Bayesian formulation}

For an introduction to Bayesian theory and inference, we refer the interested reader to \cite{Sivia2006}. 
Our goal is to construct a posterior distribution on the set of all the model parameters $\boldsymbol{\theta}$ = $\{m^1_p,m^1_c, m^2_p,m^2_c, ... p_{\rm ppE} \} $, where $m^i_p$ and $m^i_c$ correspond to the mass of the pulsar and its companion (for the $i^{th}$ observation) respectively, and $p_{\rm ppE}$ corresponds to $\alpha$ (for ppE amplitude corrections) or $\beta$ (for ppE phase corrections). According to Bayes theorem, the posterior distribution for $\boldsymbol{\theta}$ in light of measurements can be written as:
\begin{equation}
     p(\boldsymbol{\theta}|D) = \frac{{\cal L}(D|\boldsymbol{\theta})p(\boldsymbol{\theta}) }{p(D)},
     \label{eq:post}
\end{equation}
where $p(\boldsymbol{\theta}|D)$ is the probability density for the parameters $\boldsymbol{\theta}$ given the data $D$, also termed the \emph{posterior} probability density. The quantity ${\cal L}(D|\boldsymbol{\theta})$ is the \emph{likelihood} function, which represents the probability of measuring $D$ given the set of parameters $\boldsymbol{\theta}$. Finally, the quantity $p(\boldsymbol{\theta})$ is the \emph{prior} probability on $\boldsymbol{\theta}$, which represents our state of knowledge about these parameters before we analyze the data. The denominator $p(D) = \int {\cal L}(D|\boldsymbol{\theta})p(\boldsymbol{\theta}) d\boldsymbol{\theta}$ is an overall normalization constant, which is an important term to consider in model selection studies. In most parameter estimation schemes, one works with log-probability densities,
\begin{equation}
\ln p(\boldsymbol{\theta}|D) \propto \ln {\cal L}(D|\boldsymbol{\theta}) + \ln p(\boldsymbol{\theta}),
\label{eq:log-posterior}
\end{equation}
and the aim is to find the set of parameters that maximize $\ln p(\boldsymbol{\theta}|D)$. 

We now outline how we use this framework to construct the log-posterior as expressed in Eq. \eqref{eq:log-posterior}. We work with the observable $\dot{P_b}/P_b$; this is our data $D$. We assume a Gaussian model for this observable, i.e.
\begin{eqnarray}
\hspace{-0.8cm}
&\ln{\cal L}(D|\boldsymbol{\theta})& \propto -\frac{1}{2} \frac{\left(\left(\dot{P_b}/P_b  \right)_{\rm o} - \left(\dot{P_b}/P_b  \right)_{\rm th}\right)^2}{\sigma_{(\dot{P_b}/P_b)}^2}.  \label{eq: like_single}
\end{eqnarray}
Here $\left(\dot{P_b}/P_b  \right)_{\rm o}$ and $\left(\dot{P_b}/P_b  \right)_{\rm th}$ correspond to the observed value of the data and the theoretically predicted model, respectively. The latter is given by Eq. \eqref{eq:decay_phase} for phase correction, and Eq. \eqref{eq:decay_amp} for amplitude correction. Note that the above expressions are used for single measurement estimations. Obtaining joint measurement estimates is straightforward with our assumption of Gaussianity. For joint measurements, the above expressions are modified to 
\begin{eqnarray}
&{\cal L}(D^{\rm J}|\boldsymbol{\theta})& \propto \prod_{i=1}^{N} {\cal L}_i(D|\boldsymbol{\theta}) , \nonumber \\
{\rm{or}} \nonumber \\
&\ln{\cal L}(D^{\rm J}|\boldsymbol{\theta})& \propto -\frac{1}{2} \sum_{i=1}^{N} \frac{\left(\left(\dot{P_b}/P_b  \right)^{\rm o}_i - \left(\dot{P_b}/P_b  \right)^{\rm th}_i \right)^2}{\sigma_{(\dot{P_b}/P_b)_i}^2},
\end{eqnarray}
where the super-script $\rm J$ signifies \emph{joint} analysis. The $i$th individual likelihood $ {\cal L}_i(D|\boldsymbol{\theta})$ can be read from Eq. \eqref{eq: like_single}, where $i$ ranges from $1$ to $N$, $N$ being the total number of observations used for the parameter estimation. 

In the above expressions, $\sigma$ represents the error on the variable  $\dot{P_b}/P_b$ and it can be obtained by some simple error propagation. In the absence of any covariances, the uncertainty on a variable $f(x,y)$ which depends on observables $x$ and $y$ satisfies
\begin{equation*}
\sigma_{f}^2 = \left(\frac{\partial f}{\partial x}\right)^2 \sigma^2_x +  \left(\frac{\partial f}{\partial y}\right)^2 \sigma^2_y,
\end{equation*}
where $\sigma_x$ and $\sigma_y$ are the observational uncertainties on $x$ and $y$ respectively. Hence for our observable we obtain:
\begin{equation*}
\sigma_{\dot{P_b}/P_b}^2 = \left(\frac{1}{P_b^{\rm o}}\right)^2 \sigma^2_{\dot{P_b}} 
+ \left(\frac{1}{P_b^{\rm o}}\right)^2  \left( \frac{\dot{P_b^{\rm o}}}{P_b^{\rm o}}\right)^2 \sigma^2_{P_b}.
\end{equation*}

\subsection{Priors} 

Now the only piece in the puzzle left to discuss is the prior appearing on the right-hand side of Eq. \eqref{eq:log-posterior}. 
We begin by discussing the priors on the component masses, and then proceed with a discussion of our priors on the ppE parameters.  

\subsubsection*{Priors on masses}
\label{sec:priors}

We will consider three cases that differ from each other based on our assumptions about the component masses. This will allow us to study the effect of the mass prior probability on our marginalized posteriors.  For the different cases we studied in this work, the parameter space has different dimensions. For easy referencing, we tabulate them in Table \ref{tab:dimen}. Note also that, in addition to the mass-priors mentioned in this subsection, we further restrict the pulsar mass and the companion mass in the ranges: $m_p \in (0.5 , 3)~ {\rm M}_{\rm solar}$, $m_c \in (0.05,3)~ {\rm M}_{\rm solar}$.

\begin{table}[t]
	\centering
	\begin{tabular}{  c  c  c }
		\hline
		\vspace{-0.2cm}\rule{0pt}{4ex} Priors & Single & Joint\\
		\rule{0pt}{4ex}   & measurement & measurements\\
		\hline
		\hline
		\rule{0pt}{4ex} fixed-mass prior & 1 & 1 \\
		\rule{0pt}{4ex} Gaussian mass prior  & 3 & 13 \\
		\rule{0pt}{4ex} Gaussian mass-ratio prior & 3 & 9 \\		
		\rule{0pt}{4ex} Uniform mass prior & 3 & 13 \\
		\hline
	\end{tabular}
	\caption{
		Dimentionality of the parameter space for the different cases studied in this work.
	}
	\label{tab:dimen}
\end{table}

\begin{enumerate}[wide, labelwidth=!, labelindent=0pt,  label=\textbf{p\arabic*},ref=p\arabic*]
	\item \label{priors: gr} - { \emph{Priors based on GR estimates}}
	
\indent	In YH10, the authors chose the component masses of the binary to be exactly equal to the best-fit values obtained by fitting binary pulsar data to a timing model to extract ppK parameters, and then using the GR expressions for the ppK parameters as a function of the Keplerian parameters and the component masses to infer the masses. We will refer to component masses obtained in this way as \emph{best-fit masses assuming GR.} Such a choice is reasonable because no strong field observations has so far shown any evidence against GR. To be completely theory agnostic, however, one should treat the masses as free parameters to be fixed via a parameter estimation scheme. 
\vspace{-0.2cm}
	\begin{enumerate}[labelwidth=!, labelindent=0pt,  label=\textbf{p1\alph*},ref=p1\alph*]
		\item	\label{priors: gra} As a starting point (and to compare our results with those in the literature), one of our prior mass choices will be to assume that the component masses are given by their best-fit values assuming GR, i.e.~the prior on the masses are delta functions centered at the best fit values assuming GR. This prior is the same as that chosen in YH10, and we will refer to it as a \emph{fixed-mass prior}. With this prior, the parameter set reduces to $\boldsymbol{\theta}$ = $\{p_{\rm ppE} \}$, and, given a value of the ppE exponent $a$ ($b$), we have a very simple parameter estimation problem with only one parameter $\alpha$  ($\beta$) to maximize the log-posterior over. This calculation can be easily performed on a single observation, or on $N$ observations through a joint parameter estimation study.
		\item	\label{priors: grb} We will also find it convenient to relax this fixed-mass prior to study its effect on our constraints on GR. To do so, we choose a mass prior that is a Gaussian centered at the best-fit values assuming GR, with width given by the 1$\sigma$ uncertainty in the GR estimate. With this \emph{Gaussian mass prior}, for a single observation, the parameter set reduces to $\boldsymbol{\theta}$ = $\{m_{p},m_{c},p_{\rm ppE} \}$, and given a value of the ppE exponent $a$ or $b$, we then maximize the log-posterior (Eq. \eqref{eq:log-posterior}), over a three dimensional parameter space. The same calculation can then be repeated for $N$ observations through a joint analysis. 
	\end{enumerate}
	\item  \label{prior: mass ratio} - {\emph{Prior on mass ratio}}
	
\indent	 Instead of imposing a prior on each of the component masses around their GR values, we can make use of other measurements that constrain the masses. If the companion to the pulsar is bright enough for optical spectroscopy, the mass ratio $R = m_p/m_c$ can be determined through combining the Doppler shifts in the spectral lines with the timing observations of the pulsar. This is also true for double-pulsar systems for which the orbits of both NSs can be measured simultaneously. This is because for any Lorentz-invariant theory of gravity, the relative size of the orbits is related to the mass ratio of the system (up to first PN order): $R \equiv m_p/m_c = a_c/a_p$, and hence one can estimate the mass ratio of the binary.
\\ 
\indent 	One choice of prior on the masses is then to assume a Gaussian distribution on the mass ratio $R$, with width given by the 1$\sigma$ uncertainty in the measurement. We refer to this prior as a \emph{Gaussian mass-ratio prior}, and we note that out of the six observations we use as data, only four binary systems (J1012-5307, J0348-0432, J0737-3039 and J1738-0333) have separate measurements of the mass ratio, and so we only use these data for this prior. The parameter set is still $\boldsymbol{\theta}$ = $\{m_{p},m_{c},p_{\rm ppE} \}$ for a single observation, but the log posterior now depends on the log Gaussian prior on $R$ per Eq.~\eqref{eq:log-posterior}, so that only a certain region in the $\{m_{p},m_{c}\}$ parameter sub-space is favored when exploring the likelihood. 
	\item  \label{prior: uniform} -  {\emph{Uniform prior}} 

\indent	The final case we consider is to treat the masses as completely free, with uniform linear priors. This is the most ambitious case where we try to estimate ($2 N + 1 $) parameters from $N$ observations, i.e.~$N$ $m_{p}$ parameters, $N$ $m_{c}$ parameters, and 1 ppE parameter, so that  $\boldsymbol{\theta}$ = $\{m_{c,1}, m_{c,2}, \ldots,m_{c,N},m_{p,1},m_{p,2}, \ldots,m_{p,N},p_{\rm ppE} \}$. In this case, we do not expect to find reasonable constraints on all parameters, because a single ppK measurement per binary pulsar observation is not enough to estimate both component masses. As we will show later, however, even though the masses cannot be well estimated, the ppE amplitude parameter can still be constrained, albeit not as strongly as when one assumes strong priors on the masses. 
\end{enumerate}

\vspace{0.2cm}

Given these three choices of prior, which one is valid for a test of GR? As we will explain later, the deterioration of ppE constraints that occurs when we use priors (\ref{prior: mass ratio}) and (\ref{prior: uniform}) is an artifact of only using measurements of $\dot{P}_{b}$ as our data. In reality, each of the pulsars we consider also have measurements of other ppK parameters. If we were to include such measurements as data in our analysis, they would limit the range of allowed component masses to what one obtains with the Gaussian mass prior in (\ref{priors: grb}). Results obtained with this latter prior are thus the ones we quote in the introduction of this paper. Of course, we could bypass this entire discussion of priors by using (\ref{prior: uniform}) and including all measured ppK parameters. This, however, is not yet possible because a ppE mapping of these parameters requires the extension of the ppE framework to eccentric binaries; such an analysis must therefore be relegated to future work.  	

\subsubsection*{Priors on ppE corrections}
For the amplitude and phase corrections, we consider a wide range of values for the ppE exponents $a$ and $b$: $-2.8 <a < 0.8$ and $-2.0 < b< -0.6$. 
The priors on the amplitude and phase ppE parameters, $\alpha$ and $\beta$, are assumed to be \emph{uniform} in $\log \alpha$ and $ \log \beta $ respectively. The boundaries of the range are informed from YH10 estimates. We make the pessimistic assumption that the estimates in YH10 are optimistic and set our upper boundary on the ppE parameters to be slightly worse than those obtained by YH10. As an example, consider the amplitude ppE parameter when $a=-2$ for which YH10 obtains the upper limit $\alpha \lesssim 10^{-21}$. For our analysis, we would then set the prior boundaries on $\log \alpha$ to be $-40 <\log \alpha < -18$. We use similar considerations to set the boundaries for the phase correction parameter $\beta$. We then perform the optimization exercise for each discrete value of $a$ ($b$) and calculate the \emph{marginalized} 95 percentile upper limit on the ppE corrections $\alpha$ ($\beta$). We further checked that these upper limits do not change by any appreciable amount if we modify the prior limits that are set on the ppE parameter $\log \alpha$ ($ \log \beta$).

One may wonder how much the choice of the priors influences the constraints. As mentioned, in this study we have used uniform priors on $\log \rm{\theta_{ppE}}$. An alternative choice, frequently made when determining upper limits on parameters, is to use uniform priors on $\rm{\theta_{ppE}}$. Using a uniform prior on the logarithm of a parameter amounts to saying that any \emph{order of magnitude for the parameter} $\rm{\theta_{ppE}}$ is equally likely, whereas using a prior on the parameter itself amount to saying that any \emph{value of the parameter}  $\rm{\theta_{ppE}}$ is equally likely. But in practice, a uniform prior on the logarithm of a parameter will also give more weight to lower values of the parameter when exploring the posterior distribution (Eq. \eqref{eq:log-posterior}), since in this case $p(\rm{\theta_{ppE}}) \propto 1/\rm{\theta_{ppE}}$. Given that the ppE corrections are to be understood as \emph{small deformations from GR}, a uniform prior on the logarithm of the ppE parameter seems to be the appropriate choice. If one insisted on using a uniform prior on the ppE parameter itself, the bound on the ppE parameter would deteriorate by roughly one order of magnitude, leading to the largest deterioration (by a factor of 80) when $b = -0.6$.

\begin{figure}[h!]
\includegraphics[width=0.45\textwidth]{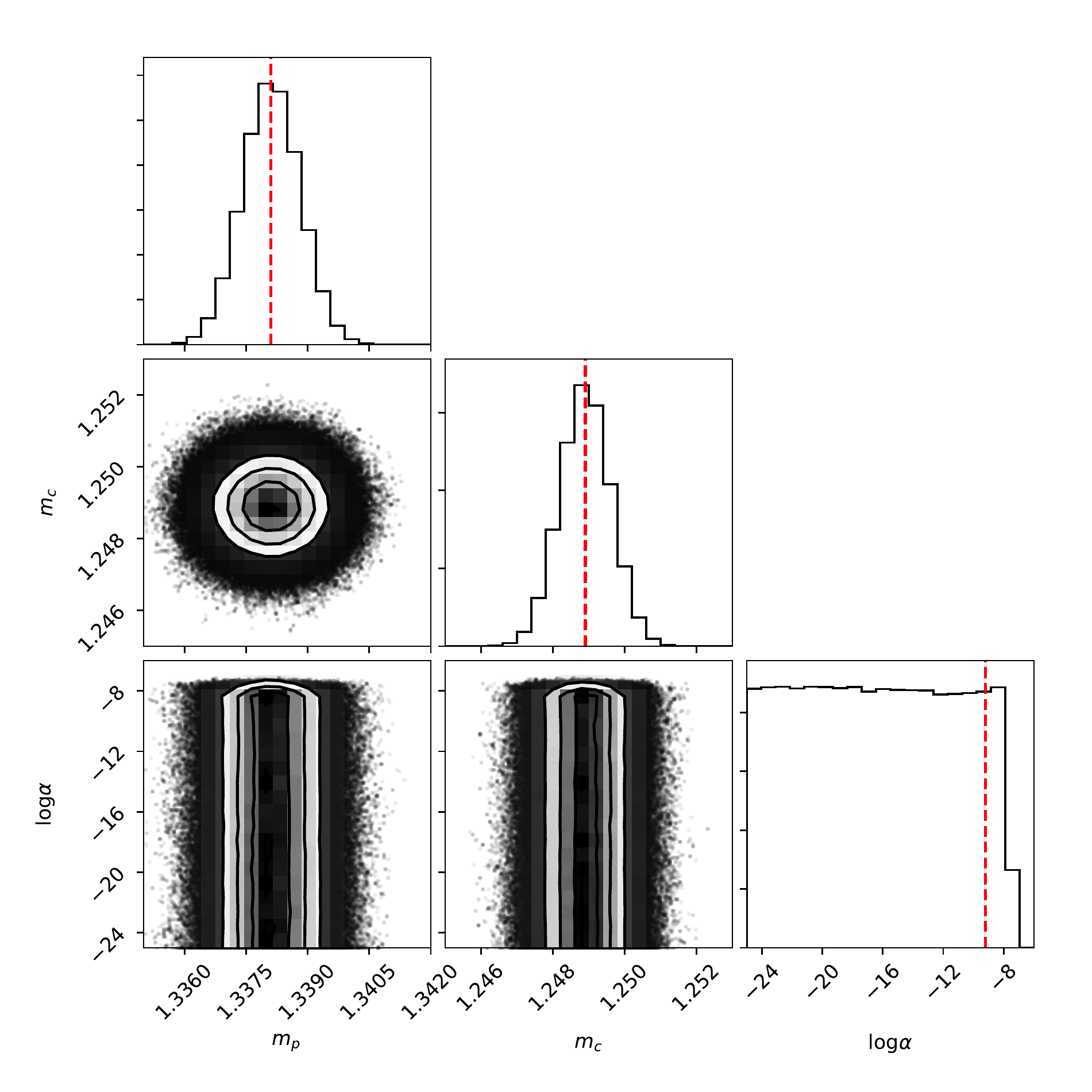} \\
\includegraphics[width=0.45\textwidth]{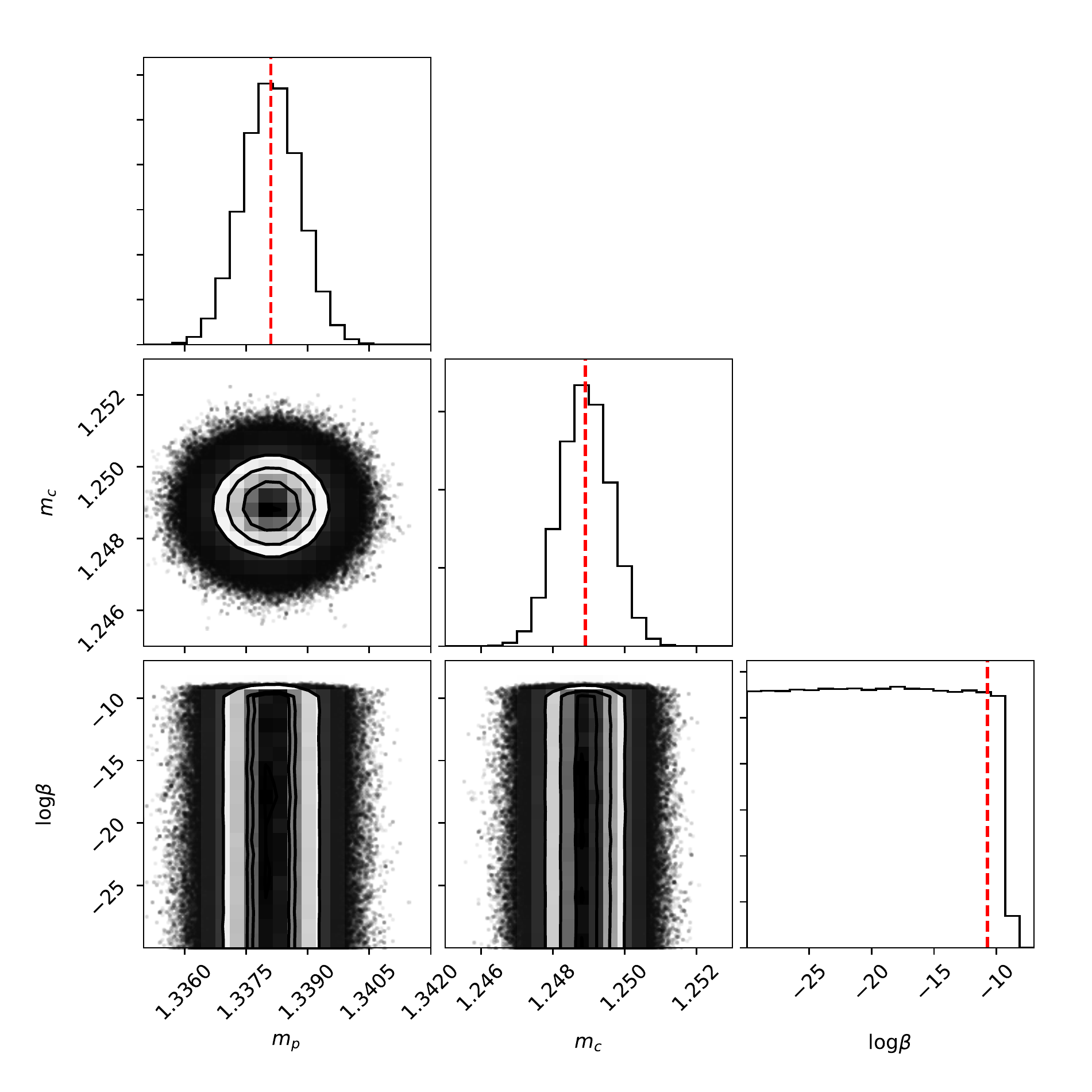}
	\caption{Constraints on the pulsar and companion masses ($m_p$ and $m_c$), and on the (log) ppE amplitude parameter $\log \alpha$ (top) and the (log) ppE phase parameter $\log \beta$ (bottom). These constraints are obtained by a Bayesian parameter estimation analysis using the measurements of PSR J0737-3039, assuming the corrections occur at $-1$PN order ($a=-2/3 $ or $b= -7/3$). We also show the mean values (`best fits') for the component masses with a dashed (red) line. For the ppE corrections (corner right plot in both panels) the dashed (red) line corresponds to the 95 percentile upper limit.}
	\label{fig:PSR3039}
\end{figure}
\subsection{Sampling the posterior}  

One can reconstruct the posterior distribution from Eq. \eqref{eq:post} in a brute-force way, by creating a simple grid and evaluating this function at each grid point to find the parameter values for which the function is maximum. It is obvious that this exercise would become computationally very expensive (almost impossible) in cases where we have a complicated likelihood function or more than a few parameters to sample. Instead, one can make use of MCMC sampling algorithms, which generate a list of samples for $\boldsymbol{\theta}$ according to the posterior distribution $p(\boldsymbol{\theta}|D)$. We use the Python package \emph{emcee} to carry out the exploration of the posterior distribution \cite{ForemanMackey:2012ig}. \emph{emcee} is an implementation of Goodman \& Weare's Affine Invariant MCMC Ensemble sampler \cite{2010CAMCS...5...65G}, which has an advantage over the standard Metropolis Hastings algorithms when the scales of covariances of the target distribution are not well known. 

\section{Results and implications}
\label{results}

In this section we present the results of our Bayesian analysis. Note that, as mentioned earlier, since we have only used one ppK measurement: $\dot{P_b} $, we can expect to find reasonable mass estimates only when we use additional mass information. This includes either using GR estimates (\ref{priors: gr}) or using measurements of the mass ratio (\ref{prior: mass ratio}). This is true even if we do a \emph {GR only} test case without any ppE corrections, because at least two ppK measurements are required to uniquely identify the component masses. 

We begin by considering constraints from a single observation, the double binary pulsar J0737-3039, so that we can compare our results to those in YH10. We then consider constraints on ppE parameters from six different binary pulsars. We conclude by computing joint constraints from the stacking of all six binary pulsar observations. 

\begin{figure*}[ht]
	\includegraphics[width=0.45\textwidth]{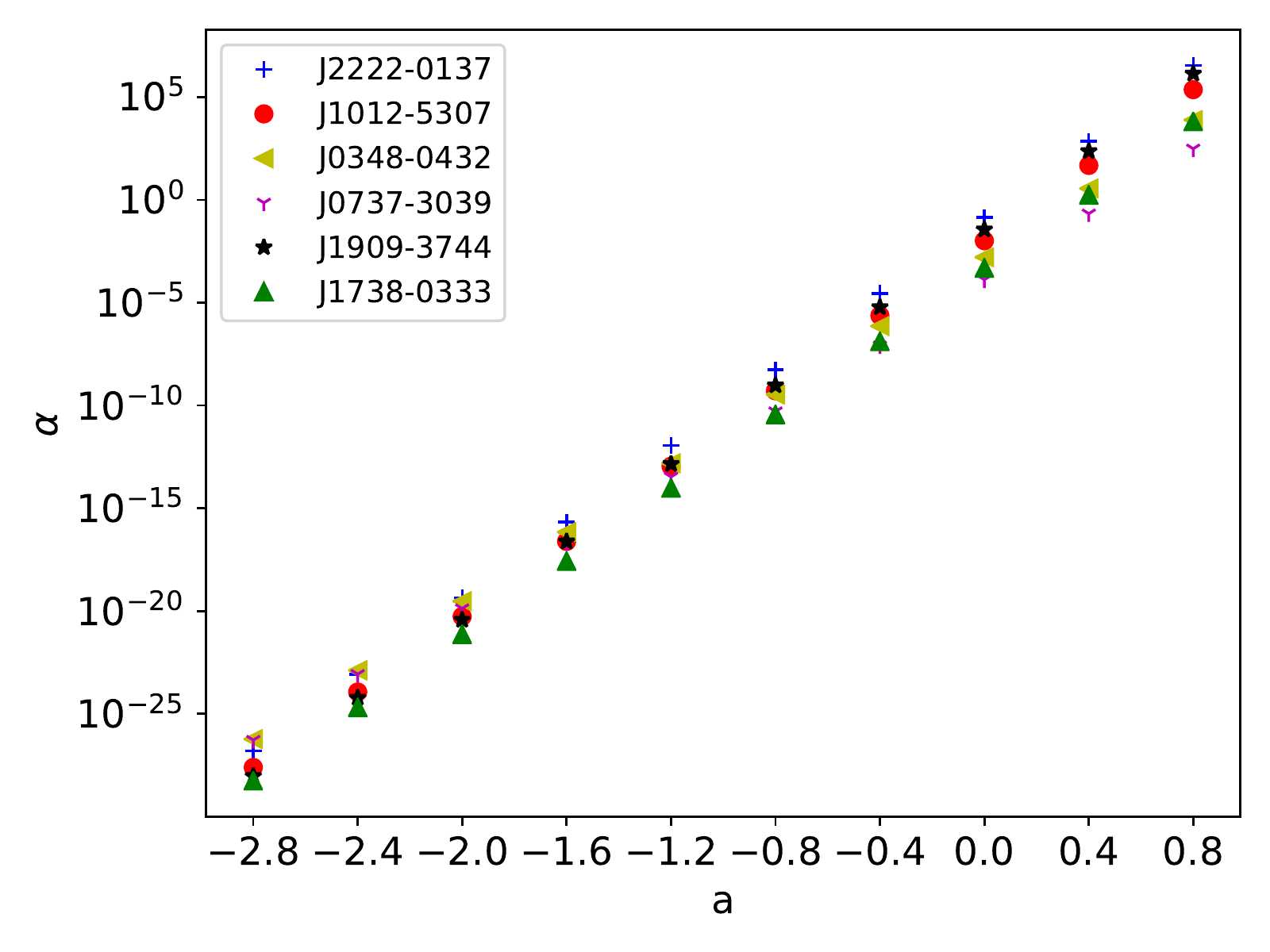}
	\includegraphics[width=0.45\textwidth]{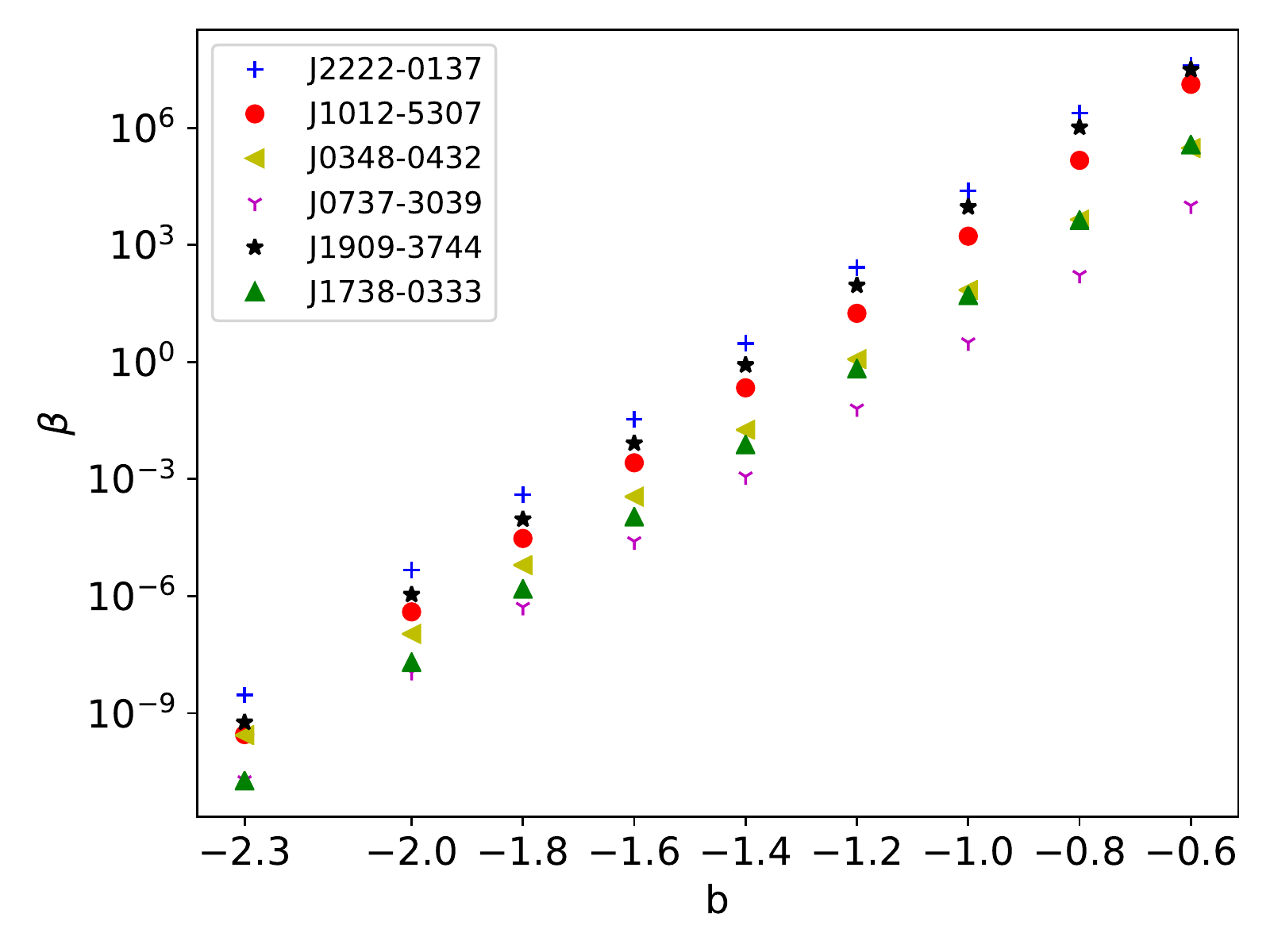}
	\caption{95 percentile upper limits on ppE corrections obtained from the different binary pulsar observations used in this work. The left and right panel correspond to amplitude and phase corrections respectively. This figure shows that the best constraints come from observations of J$1738$-$0333$ and the double pulsar J$0737$-$3039$ (for more discussion, see Sec.~\ref{subsec:indy-meas}). }
	\label{fig:ppE_individual}
\end{figure*}

\subsection{J0737-3039 and Comparisons to YH10}

We start by first comparing our results with the work of YH10, where the authors used observations for the pulsar J0737-3039, fixed the component masses to their GR values and obtained relational constraints on the ppE parameters. Our first step is then to follow their assumptions and to choose a fixed-mass prior (\ref{priors: gra}), so as to obtain the upper limit on the post-Einsteinian phase parameter $\beta$, focusing on the correction appearing at $-1$PN order as a test case. As discussed earlier, we have only one free parameter to estimate in this case: the ppE phase correction $\beta$. We find that our 95 percentile upper limits are approximately $\beta \lesssim 10^{-10}$, ($\log \beta < -10.725$) which is roughly the same as the relational constraints obtained in YH10, namely $\beta_{\rm YH10} \lesssim 10^{-10}$. We find that this agreement with YH10 extends to other PN orders as well when comparing YH10 to our 95 percentile upper limits. 

Next, we relax our assumptions on the component masses and analyze the effect of using Gaussian mass priors (\ref{priors: grb}), informed from their GR values. In this case, we need to estimate three parameters: $m_p$, $m_c$ and $\alpha$ or $\beta$. As an example case, we again focus on the ppE corrections appearing at -1PN order and show our constraints in Fig. \ref{fig:PSR3039}. The 95 percentile upper limits in this case are $\alpha \lesssim 10^{-9}$ ($\log \alpha < -9.228$) and $\beta \lesssim 10^{-10}$, ($\log \beta< -10.720$),  which is essentially the same as what we find with the fixed-mass prior. As discussed earlier, pulsar timing observations are high precision measurements and they provide very precise estimates of the component masses when multiple ppK parameters are measured. Due to the extremely small error estimates, using a Gaussian prior is almost the same as using the fixed mass prior, as is evident from the mass estimates seen in the corner plots in Fig. \ref{fig:PSR3039}. 

\subsection{Individual measurements}
\label{subsec:indy-meas}
We now study all the binary pulsars individually to see which measurements provide the best constraints on the ppE corrections using the fixed-mass prior case (\ref{priors: gra}). This choice of prior is the simplest to consider when carrying out a comparative analysis, since we only have a single parameter to estimate, but as we saw above, the resulting constraints are essentially the same as what one would obtain with a Gaussian mass prior (\ref{priors: grb}). The 95 percentile upper limits on $\alpha$ and $\beta$ are shown in Fig. \ref{fig:ppE_individual} as a function of $a$ and $b$. The two most constraining measurements are from the milli-second pulsar J$1738$-$0333$ and the double pulsar J$0737$-$3039$.

One can understand this by looking at the relation between the ppE parameters and the measurable quantities through Eq.~\eqref{eq:decay_phase} and \eqref{eq:decay_amp}. This is how YH10 originally obtained their relational constraints. If we assume that the measurements of $\dot{P_b}/P_b$ are very close to their GR values (which is a very reasonable assumption), then one can roughly relate $\alpha$ (or $\beta$) to the accuracy, $\delta$, with which these Keplerian and post-Keplerian observables are measured, namely
\begin{eqnarray}
\label{eq:rel-cons-alpha}
|\alpha| &\leq&  \left(\frac{P_b}{\dot{P_b}}\right)_{\rm GR} ~ \left(\frac{\delta}{2}\right) u^{-a}  \\
\label{eq:rel-cons-beta}
|\beta| &\leq& \left(\frac{P_b}{\dot{P_b}}\right)_{\rm GR} ~\left(\frac{5\delta}{48 |b| |b-1|}\right) u^{-b-5/3},
\end{eqnarray}
where $\delta$ is the uncertainty on the observed $(\dot{P_b}/P_b)$, while $(\dot{P_b}/P_b)_{\rm GR}$ is the GR prediction. We find that these relational constraints are the strongest for J$1738$-$0333$ and J$0737$-$3039$ because the combination $\delta \times ({P_b}/{\dot{P_b}}) \propto \delta \times r_{12}^{4} /(\eta m^{3})$ is the smallest for these binaries. That is, these two binaries are not just the two most relativistic ones, but they also have very well-measured ppK parameters. 
\begin{figure}[th]
	\begin{center}
		\hspace{-0.5cm}
	{\includegraphics[width=0.5\textwidth]{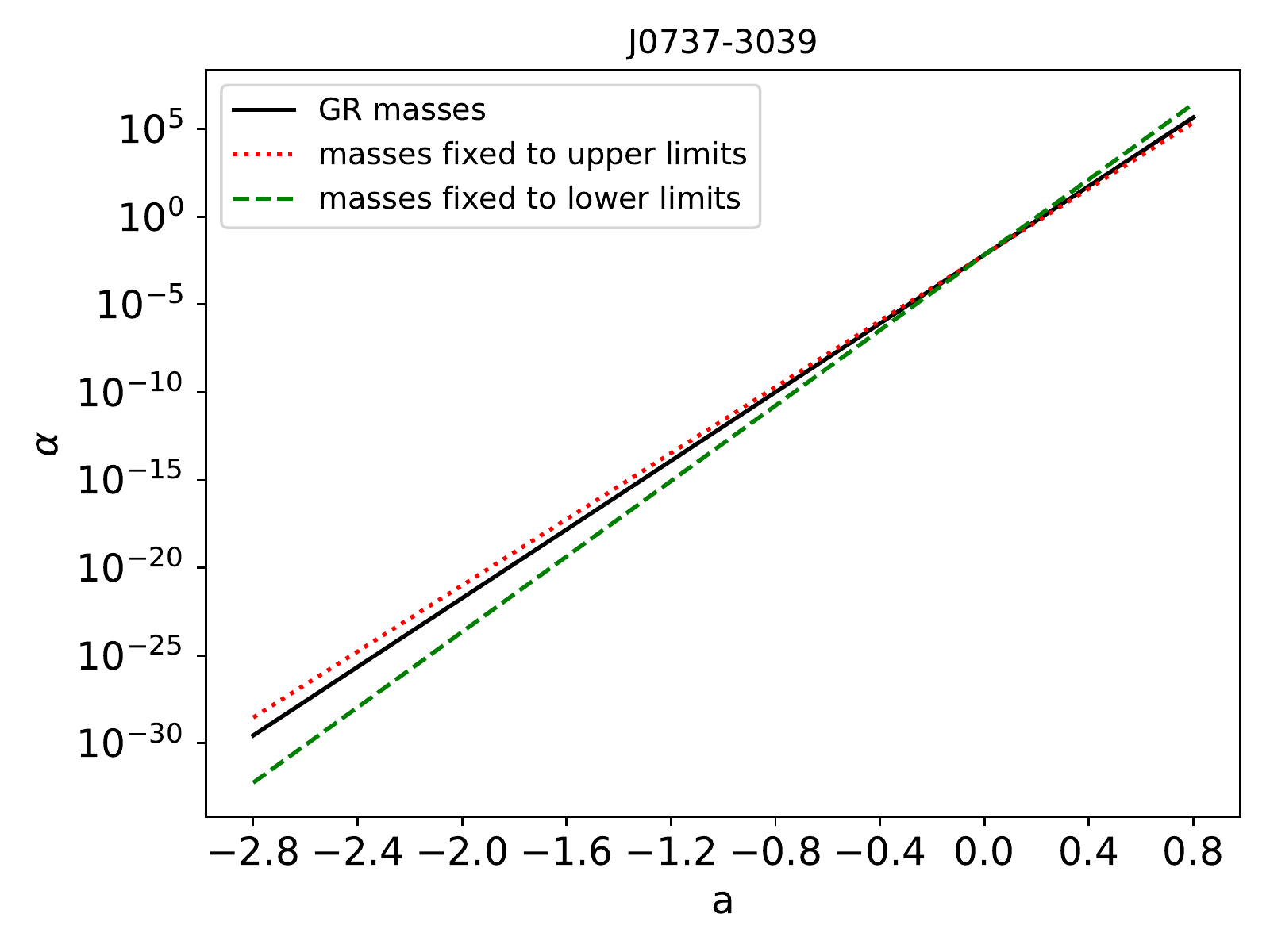}}
    \end{center}
	\caption{Constraints on the amplitude correction $\alpha$, obtained from the measurements of the binary pulsar J$0737$-$3039$. Different curves correspond to fixing the component masses to different values (see Sec \ref{subsec:indy-meas} for details). This figure shows that observations of lower mass systems can better constrain negative PN order modifications to GR. }
	\label{fig:priorEff_0737}
\end{figure}

As a further analysis, let us now study how the constraints are affected by our knowledge of the component masses. For this study, we estimate constraints on the amplitude ppE parameter while fixing the component masses to the least possible/largest possible allowed values. We show the 95 percentile upper limits obtained for one of the most informative binaries, J$0737$-$3039$, in Fig. \ref{fig:priorEff_0737}. From this figure, we observe that the binaries providing the best constraints on negative (positive) PN corrections have a lower (higher) total mass for fixed $P_{b}$ and $\dot{P}_{b}$. This can be understood by looking at the nature of the amplitude correction: $\alpha \; u^a$. For fixed frequency, lower mass values have lower values of $u$ $(= \left(\pi {\cal M}f\right)^{1/3})$, which means that if $a<0$ the ppE amplitude corrections are larger than in the high mass case. Hence, the low mass cases allow a larger modification and are therefore easier to constrain by these measurements. The situation reverses when $a>0$ which is also evident in Fig. \ref{fig:priorEff_0737}, although the effect is less pronounced since the positive $a$ range is different than the negative $a$ range in the figure. Similar conclusions can be drawn from Fig. \ref{fig:ppE_individual}. 

\subsection{Joint measurements}
\label{sec:Result-Joint}
We now use the whole set of six binary pulsar measurements to obtain joint constraints on our ppE modifications. We discuss below the results from these studies for each mass prior case considered here.

\begin{figure*}[ht]
	\begin{center}
		\begin{tabular}{clc}
			\hspace{-0.7cm}
			{\includegraphics[width=0.5\textwidth]{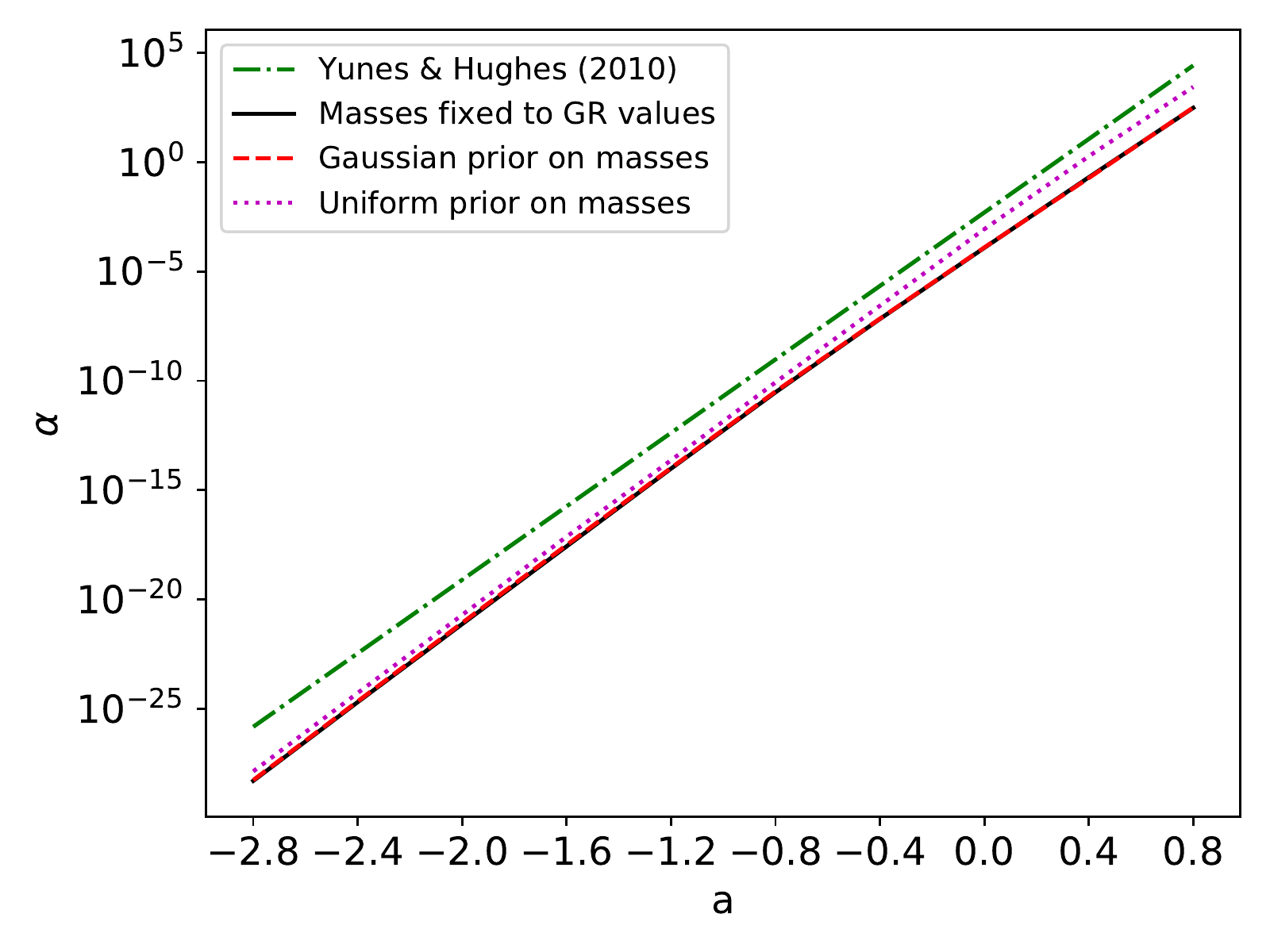}}  \quad
			{\includegraphics[width=0.5\textwidth]{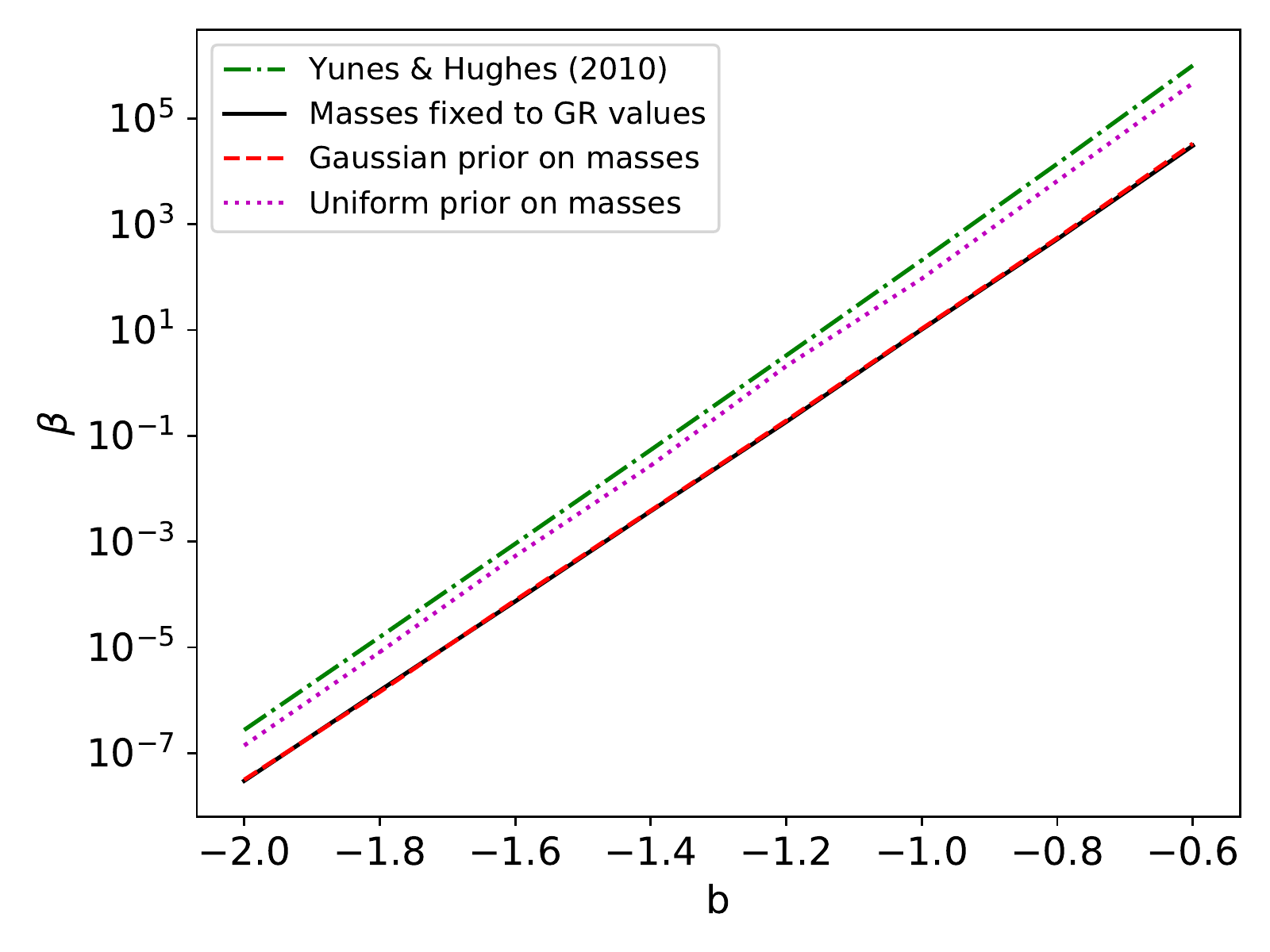}} & 
		\end{tabular}
	\end{center}
	\caption{Joint 95 percentile upper limits on the ppE amplitude parameter $\alpha$ (left) and ppE phase parameter $\beta$ (right) as a function of the ppE amplitude exponent $a$ (left) and $b$ (right), similar to Fig. \ref{fig:ppE_const_GR}. In this plot, in addition to the black curve of Fig. \ref{fig:ppE_const_GR}, we show constraints obtained using other mass priors. The figure shows that the joint constraints are about 1 or 2 orders of magnitude stronger than those found in YH10, and independent of whether one fixes the component masses (solid black curve) or uses a Gaussian prior (dashed red curve), informed from the GR estimates. The constraints deteriorate if we use uniform priors on the component masses (dotted magenta curve) because we focus on constraints derived using a single ppK measurement ($\dot{P_b}$).
 	}
	\label{fig:ppE_const}
\end{figure*}

\subsubsection{Joint results using priors based on GR estimates}
\label{sec:joint-res-GR}

The 95 percentile upper limits obtained from the joint analysis using fixed-mass priors (\ref{priors: gra}) are shown as black curves in Fig. \ref{fig:ppE_const}. Similar results hold when we use the Gaussian-mass prior (\ref{priors: grb}), shown by a red dashed curve in Fig. \ref{fig:ppE_const}. As expected, the joint constraints are better than constraints with single observations and with the relational method of YH10 by roughly 1 to 2 orders of magnitude. At all PN orders, we see improvements on the upper limits of both the amplitude and phase corrections. 
	
\begin{figure*}[ht]
	\begin{center}
		{\includegraphics[width= 0.999\textwidth, height=25cm, keepaspectratio]{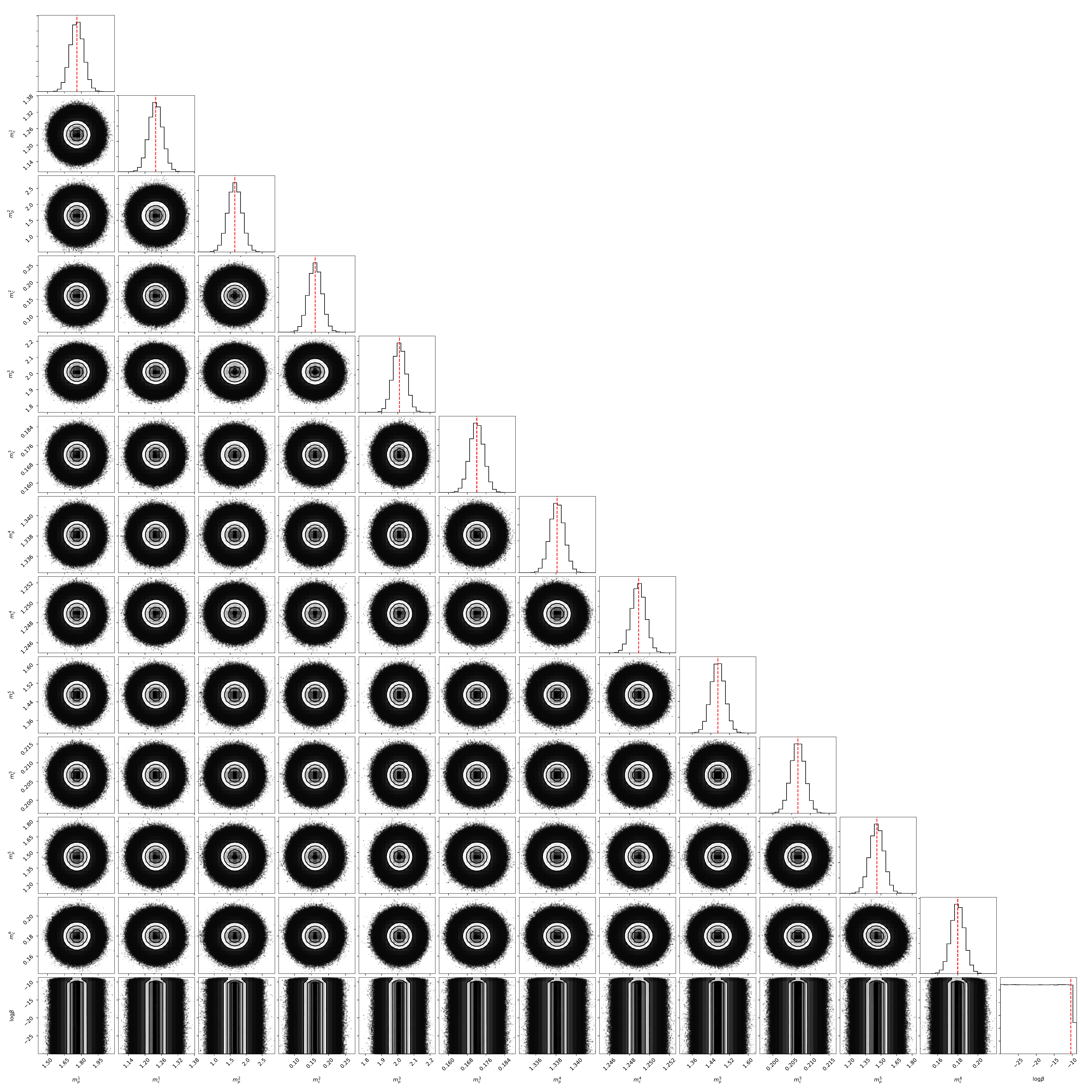}} 
	\end{center}
	\caption{Corner plot showing the joint estimation of mass parameters and the phase ppE corrections appearing at $-1$PN order. The priors on the masses are set using their GR values (Gaussian mass prior), and the prior on the ppE correction is uniform (see Sec. \ref{bayesian} for details). We also show the mean values (`best fits') for the component masses with a dashed (red) line. For the ppE corrections (corner right plot) the dashed (red) line corresponds to the 95 percentile upper limit. Observe that the estimates on the component masses are very tight owing to the use of GR estimates as priors and the the ppE correction does not appear to be correlated with any of the mass estimates. 
	}
	\label{fig:corner_beta}
\end{figure*}

As an example case of what goes into the analysis we performed, we show in Fig.~\ref{fig:corner_beta} the results of our Bayesian study for a ppE phase correction at -1PN order ($b=-7/3$) using a Gaussian-mass prior. We do not observe any significant changes in the corner plot when we look at ppE amplitude corrections, or ppE corrections at other PN orders. The corner plot shows that the masses do not present significant correlations, with each marginalized posterior considerably Gaussian. Again, this is due to the very high precision with which the masses are estimated by the pulsar timing analysis. The construction of Fig.~\ref{fig:ppE_const} required the calculation of corner plots like that shown in Fig.~\ref{fig:corner_beta} at each PN order sampled. 

\subsubsection{Joint results using priors on the mass ratio}

	 As an example of how the marginalized posteriors changes when we use the Gaussian mass-ratio prior, Fig. \ref{fig:corner_dat6} shows the corner plot for a ppE phase correction at -1PN order ($b=-7/3$), for the binary J1738-0333. In this figure we compare the Gaussian mass prior (black contours) to the Gaussian mass-ratio prior (gray contours), and we observe that the marginalized posteriors on the component masses have a larger uncertainty than when we use GR informed priors on the masses. This widening in the joint posterior is responsible, in part, for the slight deterioration of the ppE constraint.  Similar results are obtained for all four binaries for which mass-ratio measurements are available (see Sec.~\ref{sec:priors}). 
	
\begin{figure*}[ht]
	\begin{center}
		\begin{tabular}{clc}
			\hspace{-0.7cm}
			{\includegraphics[width=0.5\textwidth]{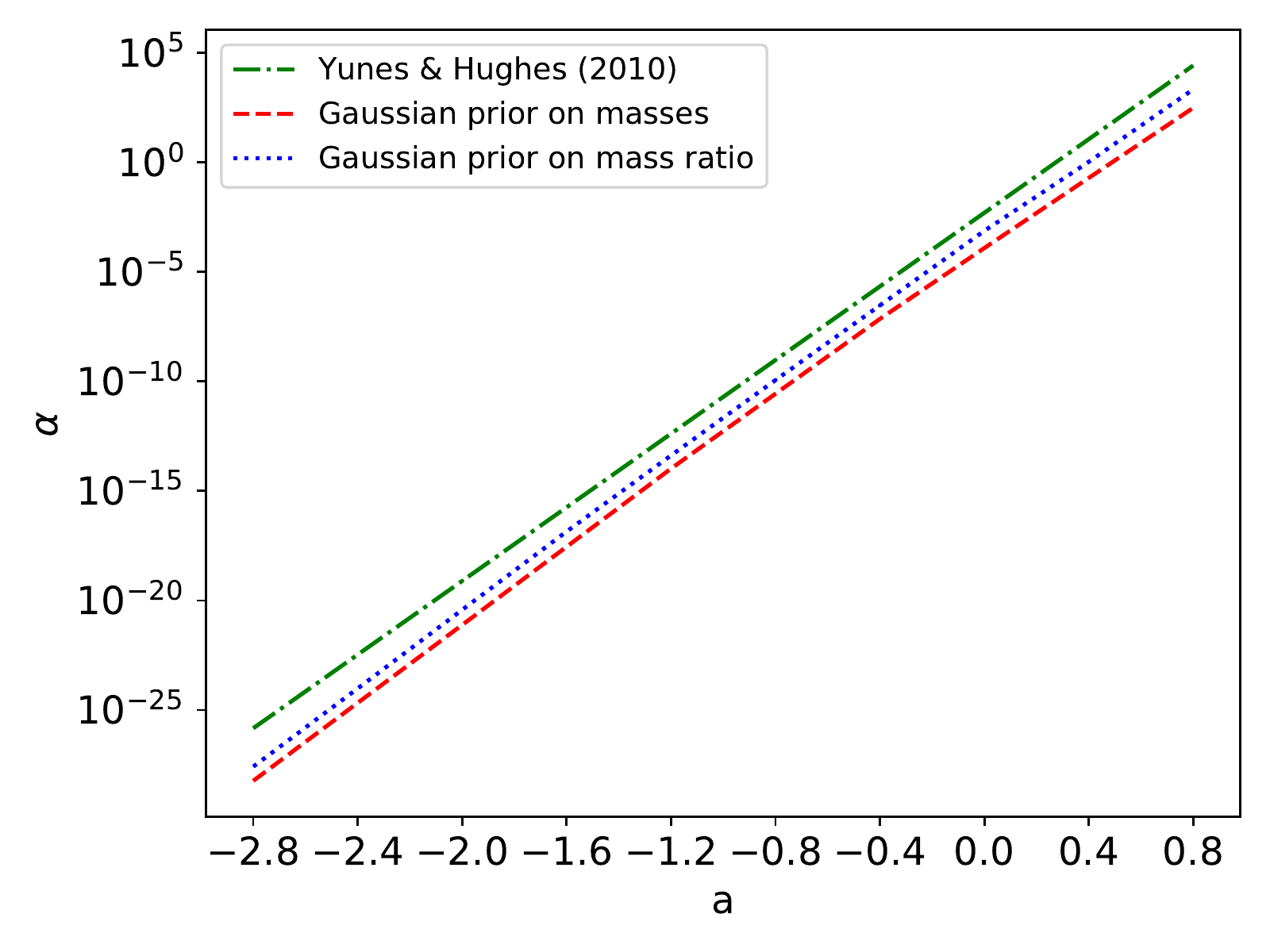}}  \quad
			{\includegraphics[width=0.5\textwidth]{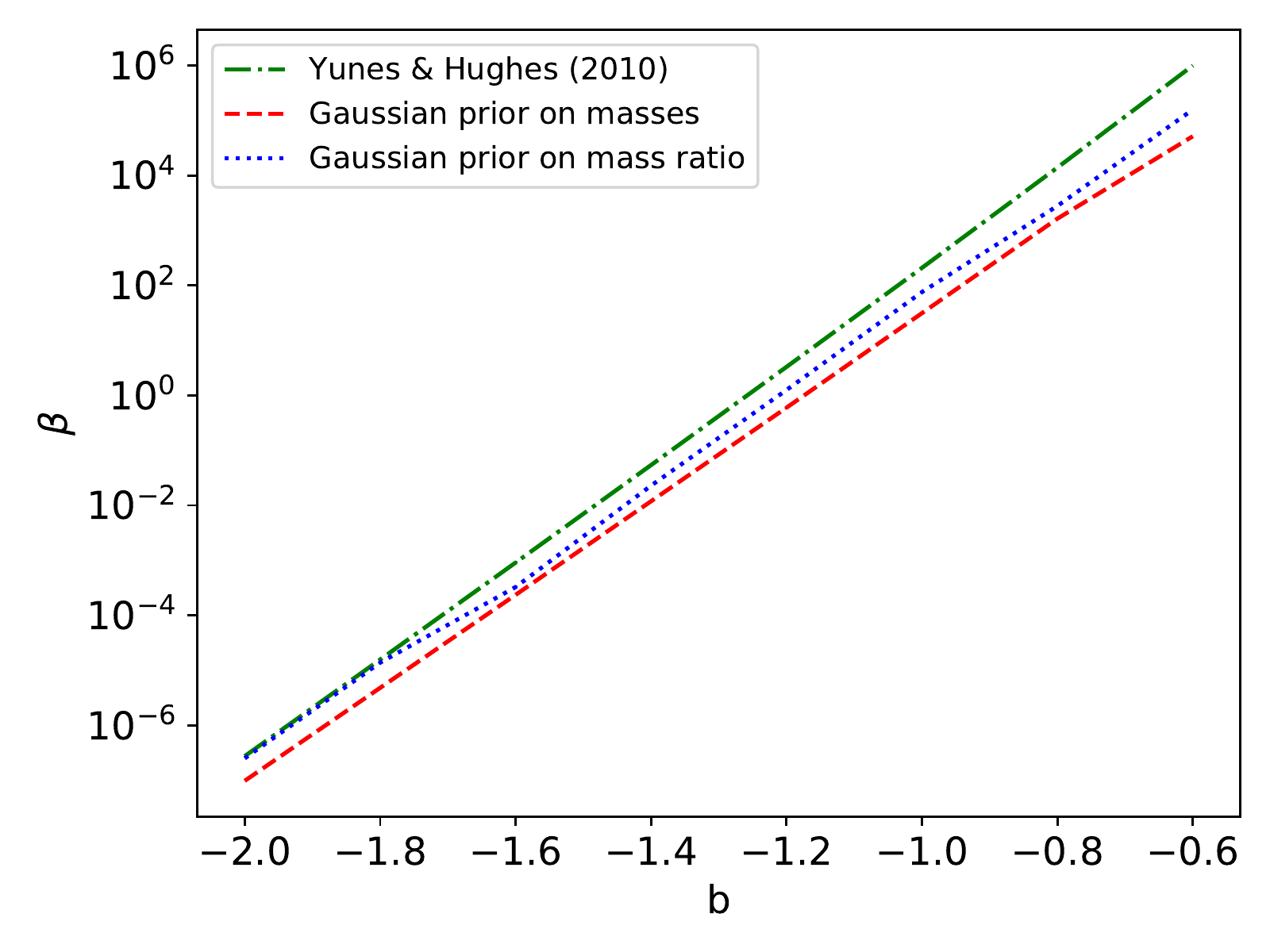}} & 
		\end{tabular}
	\end{center}
	\caption{Joint 95 percentile upper limits on the ppE amplitude parameter $\alpha$ (left) and ppE phase parameter $\beta$ (right) as a function of the ppE amplitude exponent $a$ (left) and $b$ (right), similar to Fig. \ref{fig:ppE_const_GR}. In this plot, however we compare constraints obtained using only the four binaries for which mass-ratio measurements are available. The figure shows that the joint constraints worsen if we use a prior on the mass ratio (dotted blue curve) instead of priors on the component masses derived assuming GR (dashed red curve).
	}
	\label{fig:ppE_const_R}
\end{figure*}

\begin{figure*}[ht]
	\begin{center}
		{\includegraphics[width= 0.99\textwidth, height=15cm, keepaspectratio]{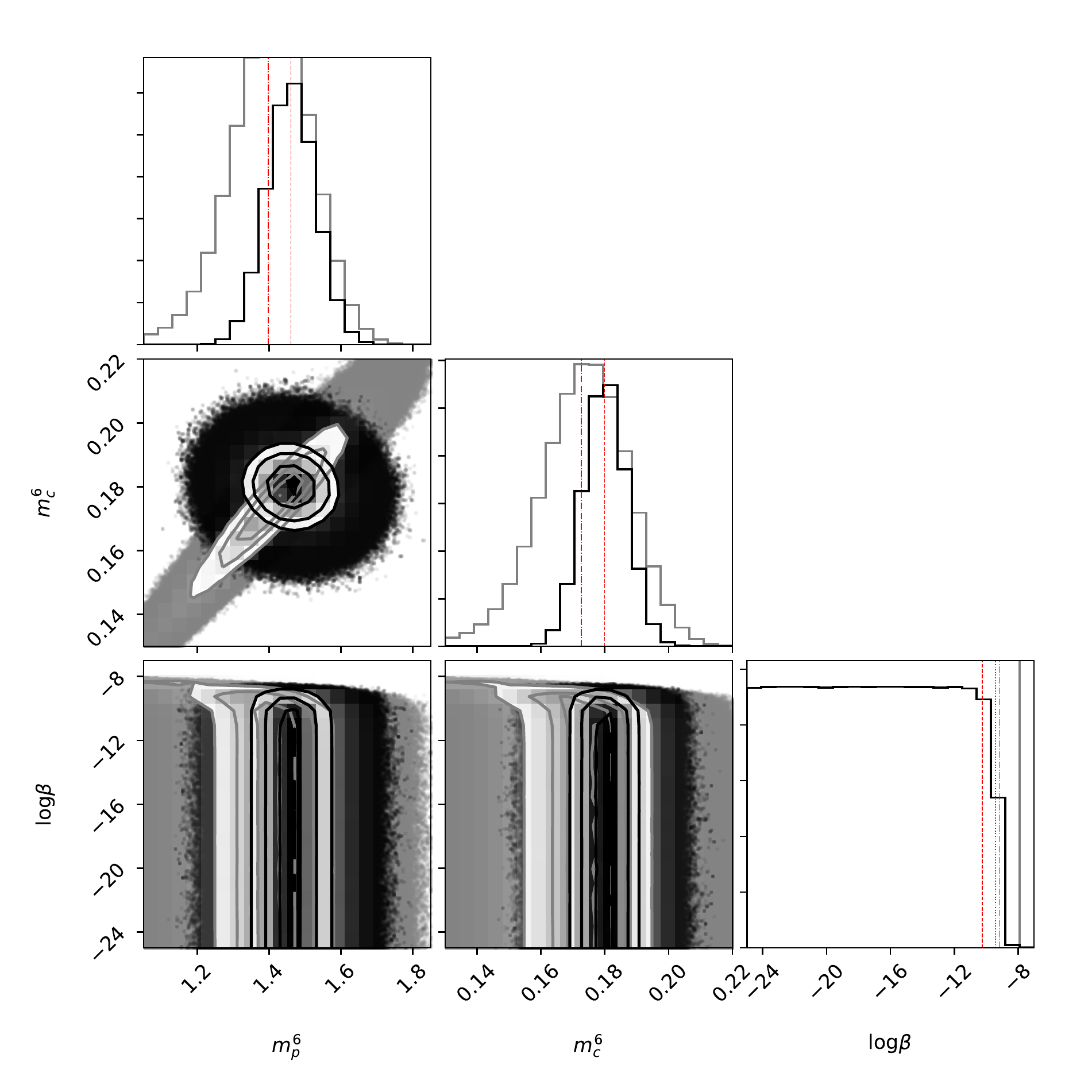}} 
	\caption{ Corner plot showing the joint estimation of mass parameters and the phase ppE corrections appearing at $-1$PN order, for the binary J1738-0333 . Black contours correspond to the case where the priors on the masses are set using their GR values (Gaussian mass prior), and the prior on the ppE correction is uniform. The gray contours, shown here for comparison, correspond to the case where the measurement of the mass ratio is used as a prior. We also show the mean values (`best fits') for the component masses with a dashed (dash-dotted) line for the Gaussian mass prior (mass ratio prior) case. For the ppE corrections (corner right plots) the dashed (dash-dotted) line corresponds to the 95 percentile upper limit for the two cases. Observe that the estimates on the component masses worsen when we use the mass-ratio prior, compared to the Gaussian mass prior case, but the effect on the estimate of the ppE correction is mild. We also show the upper limit obtained when using uniform priors on masses (dotted red line), but corresponding histograms for the masses are not shown since we only recover the priors on the masses in this case. }
	\label{fig:corner_dat6}
	\end{center}
\end{figure*}

	Next we do the joint analysis using all four binaries. The 95 percentile upper limits using the Gaussian mass-ratio prior (\ref{prior: mass ratio}) are shown as dotted blue curves in Fig.~\ref{fig:ppE_const_R}. We also show constraints obtained by analyzing these four binary observation with Gaussian mass prior (\ref{priors: grb}) (as opposed to six observations in Sec.~\ref{sec:joint-res-GR}). Again, as expected, we see the constraints slightly worsen, at all PN orders, if we do not include information from the GR estimates (dashed red curve). This is because lifting the prior on the masses allows the chains to explore a wider region in the component-masses subspace, and since we are only using 1 ppK parameter here, the masses are not well constrained. This inflates the marginalized posterior on the ppE deformation deteriorating the bound.
	
	We refrain from plotting a curve corresponding to this case in Fig. \ref{fig:ppE_const} since the number of observations analyzed in this case is different from all other cases we considered (similar curves are shown in Fig. \ref{fig:ppE_const_R} instead). Nevertheless, If one were to plot it, the curve will closely trace the dotted magenta line corresponding to the uniform mass prior case.

\subsubsection{Joint results using a uniform mass prior}
\label{res:uni} 

If we use flat priors on the masses, without any additional information (\ref{prior: uniform}), our overall estimation of the parameters worsens. We find that we cannot estimate the component masses for any of the six binaries with significant confidence. Again, this is due to the fact that we are using a single ppK measurement for the analysis without any additional information on the masses. Hence, we have to explore a much larger parameter space with very limited information. The 95 percentile upper limits on the amplitude and phase ppE corrections, at various PN orders, are shown as dotted-magenta curves in Fig. \ref{fig:ppE_const}. We find that the constraints on the phase corrections are now around 1--2 orders of magnitude weaker than the case in which the mass priors are chosen based on GR estimates. 
	
	To present an example of how our estimates change when we use this prior, Fig. \ref{fig:corner_dat6} shows the 95 percentile upper limit for the ppE phase correction at -1PN order ($b=-7/3$), estimated by analysing the binary J1738-0333. We do not show the full corner plot for this case as the marginalized histograms for the component masses will merely show that we manage to recover the priors. In spite of this, we find that, although removing prior information about the component masses has a disastrous effect on the estimation of the masses, it does not prevent us from constraining the ppE parameters. Clearly, there is a deterioration of the ppE constraints by an order of magnitude, but a constraint is still possible, even though the masses cannot be estimated. The reason for this is that a flat prior still has a boundary, so the chains are not allowed to explore \emph{any} value of the component mass subspace. This limitation is apparently sufficient to allow for ppE constraints. 

Given that the ppE constraints depend on the priors, as shown for a test case in Table \ref{tab:comp_priors}, which bound should we take seriously? The answer to this question reveals itself when we understand \emph{why} the ppE constraints depend on the priors. As we have explained throughout this paper, this is because when one uses a single ppK parameter, the component masses of the binary pulsar cannot be measured, due to the strong degeneracy between these parameters (most evident in Fig. \ref{fig:corner_dat6}). This induces a widening of the marginalized posterior on the ppE constraints. In that sense, this variation in the ppE estimates for different mass priors is artificial and just a consequence of working with a single ppK measurement. If we instead had worked with two ppK measurements (such as $\dot{P}_{b}$ and the Einstein delay) or more ppK parameters, then there would be enough information in the data to constrain the component masses much better, yielding ppE constraints that are similar to what we find when we use Gaussian mass priors. An extension of this paper to include multiple ppK parameters is left to future work, as it first necessitates the extension of the ppE framework to eccentric binaries.

%
\begin{table}[t]
	\centering
	\begin{tabular}{ c  c  c  c }
		\hline
		\rule{0pt}{4ex} $m_{\rm GR}$ & ~$N(m^{\rm GR},\sigma^{\rm GR}_m)$ & ~$N(R,\sigma_R)$ & ~Uniform in mass\\
		\hline
		\hline
		\rule{0pt}{4ex} $-10.3$ & $-10.2$ & $-9.7$ & $-9.8$\\
		\hline
	\end{tabular}
	\caption{
		Comparison between the 95 percentile upper limits obtained on the magnitude of the phase ppE correction $\log \beta$ appearing at -1 PN order i.e. for $b=-7/3$ (Eq. \eqref{eq: h_ppE}) for the different cases studied in this paper. 
	}
	\label{tab:comp_priors}
\end{table}

Let us then close by providing an analytic fit for the constraints on the amplitude and phase ppE constraints as a function of the ppE exponents in the case of Gaussian mass priors. It is easiest to fit $\log \alpha$ ($\log \beta$) to $a$ ($b$) since the relationship is very close to {\emph{linear}}, as is demonstrated in Fig.~\ref{fig:ppE_const_GR}, and also in YH10. We obtained the following fitting functions from our constraints for the fixed mass case:
\begin{eqnarray}
\log \alpha &=& -3.9822 ~a + 8.5753,  \\
\log \beta &=& 9.5777~ b + 8.5618,
\end{eqnarray}
where we have used the 95 percentile upper limits.

\subsection{Comparison with LVC constraints}

The LIGO-Virgo collaboration (LVC) has also performed GR tests and released constraints on model-independent deviations from GR ~\cite{LIGOScientific:2019fpa}. In these studies relative shifts in the PN coefficients of the Fourier phase, $\delta \phi$ were constrained using GW events, where $\delta \phi$ is defined using ($i$ corresponds to non-GR corrections entering at different PN orders):
\begin{equation}
\phi_i = \phi_i (1+ \delta \phi_i),
\end{equation}
and then treated as an additional free parameter in the parameter estimation scheme. Note that in the LVC paper $i=n$ corresponds
to $(n/2)$PN order and we follow this scheme here. 

To compare binary pulsar constraints to those obtained by the LVC, we perform a study similar to that of the previous subsections, but now treating
$\delta \phi_i$ as our GR correction parameter instead of using the ppE phase correction $\beta$. We use flat priors in $\delta \phi_i$, and obtain posterior samples for the corrections to compare the upper limits with LVC constraints. This exercise is straightforward 
for the PN corrections appearing at -1PN, -0.5PN, 0PN and 0.5PN order, since $\delta \phi$ can be very easily related to $\beta$ for
these cases \cite{Yunes:2016jcc}:
\begin{eqnarray}
&\beta_{\rm -1PN} &= \frac{3}{128} \delta \phi_{-2} \eta^{2/5}, \nonumber \\
&\beta_{\rm -0.5PN} &= \frac{3}{128} \delta \phi_{-1} \eta^{1/5},  \nonumber  \\
&\beta_{\rm 0PN} &= \frac{3}{128} \delta \phi_{0},  \nonumber  \\
&\beta_{\rm 0.5PN} &= \frac{3}{128} \delta \phi_{1} \eta^{-1/5}.
\end{eqnarray}
For other PN corrections, one would have to account for additional physical parameters like spin, which is beyond the scope of this paper. 

Binary pulsar constraints on the $-1$PN, $0$PN and $0.5$PN corrections can be directly compared to those obtained by
the LVC. Since the LVC did not release the posterior samples for combined measurements, we use the posterior samples 
obtained from the most constraining GW event, GW170608, for our comparison~\cite{LIGOposteriors}. These constraints
are very close to the joint LVC constraints, as is evident from Fig. 4 in~\cite{LIGOScientific:2019fpa}. 

The comparison
between our binary pulsar constraints and those obtained by LVC is shown in Table \ref{tab:comp_LVC}. We see that the constraints obtained 
from binary pulsar measurements are competitive with those obtained from GW measurements at 0PN order, and much tighter 
than LVC constraints at -1PN correction, as expected. But LVC constraints become tighter than binary pulsar ones at positive PN order. 
Hence our constraints at lower (negative) PN orders can be used as informed priors for future LVC studies.
\begin{table}[t]
	\centering
	\begin{tabular}{ c  | c  c  }
		\hline
		\rule{0pt}{4ex} PN order & ~LVC & ~ This work\\
		\hline
		\hline
		\rule{0pt}{4ex} -1PN & $5.4 \times 10^{-3}$ & $10^{-7}$ \\ 
		\rule{0pt}{4ex} 0PN & $8.9 \times 10^{-2}$ & $2.5 \times 10^{-2}$ \\
		\rule{0pt}{4ex} 0.5PN & $1.9 \times 10^{-1}$ & $1.6 \times 10^1$ \\		
		\hline
	\end{tabular}
	\caption{
		Comparison of our 95 percentile upper limits on $\delta \phi$ using joint measurements of binary pulsars, with those obtained by LVC ~\cite{LIGOScientific:2019fpa,LIGOposteriors} using the measurements from GW170608.
	}
	\label{tab:comp_LVC}
\end{table}

\section{Implications} 
In this work we have presented constraints on the ppE framework using Bayesian analysis of six binary pulsar observations. The constraints, which are 95 percentile upper limits, are 1--2 orders of magnitude tighter than those obtained through approximate (non-Bayesian) methods with the double binary pulsar in YH10. We have further found that our constraints on the ppE corrections are robust to our assumptions regarding the component masses, i.e. different mass priors.

Our most important results are shown in Fig. \ref{fig:ppE_const}. The areas above the black solid curves are excluded by the joint measurement of six binary pulsar observations. This implies that the amplitude and phase correction magnitudes, $\alpha$ and $\beta$, have to be smaller than these upper limits at the different PN orders shown in the figure. The general trend in these plots, of increasing upper limits with increasing values of $a$ or $b$, is as expected from theoretical considerations (see e.g. YH10). This can be understood by looking at the mathematical structure of the ppE corrections (see e.g.~Eq. \eqref{eq: h_ppE}). The ppE corrections become smaller the larger the value of $a$ or $b$, since the systems we are studying are low velocity sources. We also showed that for constant $P_b$ and $\dot{P_b}$ (and the same measurement accuracy), binary systems with lower (higher) total mass give tighter upper limits on negative (positive) PN ppE corrections (see e.g. Fig. \ref{fig:priorEff_0737}). Our constraint on the generic GR deviation parameter at -1PN order, $\delta \phi_{-2}$, is around 4 orders of magnitude tighter than the corresponding LVC constraint as of the submission of this manuscript.

This is the first study of ppE constraints with binary pulsar observations that analyzes assumptions regarding the component masses We analyzed different mass priors (\ref{sec:priors}) and found that the major contribution in determining these upper limits is the number of binary pulsar observations, how relativistic these binaries are, how well the ppK parameters are determined, and how many ppK measurements are used for each of these observations. We also studied how much these constraints could vary if we assumed uniform priors on the masses. Since we used a single ppK measurement for these binaries, in the absence of any additional mass information, we obtained un-informative posteriors on masses, but the effect on the estimates of the ppE corrections was considerably milder, suggesting the constraints derived here are robust.

Our ppE constraints can be used as priors when performing similar studies using other pulsar observations or future GW observations. An interesting extension to this project, which is ongoing, is to relate other post-Keplerian parameters to ppE corrections so that more data from binary pulsar observations can be utilized for testing GR. This task is complicated because it may require a deeper understanding of how to extend the ppE framework to more generic, eccentric and possibly spinning binaries. An interesting approach would be to carry out an effective field theory treatment in which a ppE-corrected Einstein-Infeld-Hoffman Lagrangian is derived to obtain ppE-corrected (conservative) equations of motion. The latter would be important to map ppK parameters to ppE parameters.  

\acknowledgments
We thank David Anderson for discussions and Neil Cornish for constructive comments on the manuscript. We also thank the anonymous referee for making many useful suggestions. This work was supported by
NSF grant PHY-1759615, and NASA grants 80NSSC18K1352,  NNX16AB98G and 80NSSC17M0041.

\bibliographystyle{apsrev4-1}
\bibliography{biblio}

\end{document}